# ATOMIC TRANSITION PROBABILITIES FOR UV AND BLUE LINES OF Fe II AND ABUNDANCE DETERMINATIONS IN THE PHOTOSPHERES OF THE SUN AND METAL-POOR STAR HD 84937

(Short title: ATOMIC TRANSITION PROBABILITIES OF Fe II)

E. A. Den Hartog[1], J. E. Lawler[1], C. Sneden[2], J. J. Cowan[3], and A. Brukhovesky[1]

[1]Department of Physics, University of Wisconsin-Madison, 1150 University Ave, Madison, WI 53706; eadenhar@wisc.edu; jelawler@wisc.edu; abrukhovetsk@wisc.edu

[2]Department of Astronomy and McDonald Observatory, University of Texas, Austin, TX 78712; chris@verdi.as.utexas.edu

[3]Homer L. Dodge Department of Physics and Astronomy, University of Oklahoma, Norman, OK 73019; jjcowan1@ou.edu

Abstract

We report new branching fractions for 121 UV lines from the low-lying odd-parity levels of Fe II belonging to the $z^6D^o$, $z^6F^o$, $z^6P^o$, $z^4F^o$, $z^4D^o$ and $z^4P^o$ terms of the $3d^6(^5D)4p$ configuration. These lines range in wavelength from 2250 – 3280 Å and originate in levels ranging in energy from 38459 – 47626 $cm^{-1}$. In addition, we report branching fractions for 10 weak blue lines connecting to the $z^4D^o$ term which range in wavelength from 4173 – 4584 Å. The BFs are combined with radiative lifetimes from the literature to determine transition probabilities and log(*gf*) values. Comparison is made to selected experimental and theoretical data from the literature. Our new data are applied to iron abundance determinations in the Sun and in metal-poor star HD 84937. For the Sun, eight blue lines yield log ε(Fe) = 7.46 ± 0.03, in agreement with standard solar abundance estimates. For HD 84937 the observable wavelength range extends to the vacuum UV (λ ≥ 2327 Å), and from 75 lines we derive log ε(Fe) = 5.26 ± 0.01 (σ = 0.07), near to the metallicity estimates of past HD 84937 studies.

# 1. INTRODUCTION

The iron-group (Fe-group, 21≤ Z≤30) elements in the oldest metal-poor (MP) stars were produced in supernovae of the first massive stars early in the history of the Galaxy. As such, the abundance patterns of these stars represent a fossil record of these early explosive events. Models of these early supernovae can be compared to observational studies of the relative Fe-group abundance trends with metallicity to further the understanding of Fe-group nucleosynthesis. However, model predictions do not always match observation. (e.g. Henry et al. 2010; Sneden et al. 2016) If the observations are to be useful to improve the models, it is imperative that the observed abundances are known to be reliable. The current study is part of a larger, ongoing effort by our collaboration to systematically improve the reliability of Fe-group abundance determinations in MP stars.

Our methodology is to improve the laboratory data for each Fe-group specie and apply these new data to determine both accurate and precise stellar abundances for the Fe-group elements over a range of stellar metallicities. We explore the limits of local thermodynamic equilibrium (LTE) assumptions for that element in metal-poor stars in order to discriminate against lines that would be poor abundance indicators. The bulk of Fe-group elements in the line-forming layers of F, G, and K stars, which are of interest in abundance studies extending from Solar metallicity stars to extremely MP stars, is singly-ionized (e.g., Sneden et al. 2016). Transitions which probe the ground and low-lying metastable levels of the singly-ionized Fe-group species cannot be seriously out of equilibrium. But lines that probe the neutral species, particularly the resonance lines of the neutral, are sometimes found to be unreliable abundance indicators. Earlier studies by our collaboration found this to be the case for both Mn I (Sneden et al. 2016) and Cr I (Lawler et al. 2017) resonance lines.

Some studies are turning to non-LTE modeling (e.g. Bergemann & Gehren 2008; Bergemann & Cescutti 2010, Bergemann et al. 2017), but such models are still hindered by the lack of reliable cross-sections and rate constants, particularly for inelastic and super-elastic collisions of H and He atoms with metal atoms and atomic ions. The situation is improving – Barklem is making significant contributions to the computation of better cross sections and rate constants for inelastic and superelastic heavy particle collisions (see, e.g., Barklem & Aspelund-Johansson 2005; Barklem et al. 2011; Barklem 2016) to replace the widely used Drawin approximation [Drawin 1968, 1969]. Accurate rate constants for other important reactions will also be needed for reliable non-LTE modeling. There are also effects from convection beyond those covered by a single microturbulance parameter in one dimensional (1D) models. The development of full three dimensional photospheric models including a rate equation treatment of NLTE is being undertaken by some groups (Asplund 2005; Asplund et al. 2009). In the present series of papers, we have instead relied on LTE(1D) modeling with careful choice of transitions for abundance probes.

Our method for determining transition probabilities (A-values) is to combine branching fractions (BFs) determined from high-resolution emission spectra with radiative lifetimes measured using time-resolved laser-induced fluorescence. The A-values are then converted to log(*gf*)s (the logarithm of the absorption oscillator strength multiplied by the degeneracy of the lower level of the transition). Our current effort in the Fe-group started with Cr I (Sobeck et al. 2007) and then

Mn I and Mn II (Den Hartog et al. 2011). Because of the relatively high abundances of Fe-group species, we have tried to focus on measurement of weaker branches for determining abundances in stars with metallicities near that of the Sun. These transitions will often lie on the linear part of the curve of growth (hereafter c-o-g), and will therefore be more reliable abundance indicators.

In past studies we have primarily used spectra from Fourier Transform Spectrometers (FTSs) for the determination of branching fractions. FTSs have many advantages for branching fraction work but they are not ideal for measuring weak branches due to the multiplex noise inherent in all interferometric spectrometers. Multiplex noise results from the smooth redistribution across an interferogram of the Poisson statistical noise from all lines in the spectrum. This results in the weak branches receding into the noise as the source current is made low enough to have the strong branches optically thin. The desire to measure weak branching fractions has led to the development of the University of Wisconsin (UW) high-resolution, 3-m focal length echelle spectrometer. This instrument has high resolving power (>250,000), broad spectral coverage and excellent UV sensitivity (Wood & Lawler 2012). Using a combination of FTS data and 3-m echelle data, progress has been made on: Ti I (Lawler et al. 2013); Ti II (Wood et al. 2013); V II (Den Hartog et al. 2014a; Wood et al. 2014a); Ni I (Wood et al. 2014b); V I (Lawler et al. 2014; Wood et al. 2018); Fe I (Den Hartog et al. 2014b; Ruffoni et al. 2014, Belmonte et al. 2017); Co I (Lawler et al. 2015); Cr II (Lawler et al. 2017) and Co II (Lawler et al. 2018).

The abundance peak of the Fe-group is, of course, at iron. The high abundance of iron in astrophysical objects along with its rich spectrum makes it one of the most important elements in stellar astrophysics. Because most of the iron is singly-ionized in MP stars of interest, Fe II is a spectrum of considerable interest, and has therefore been the subject of a number of laboratory studies to improve and expand the available transition probability data. A summary of all this earlier work on Fe II transition probabilities is beyond the scope of this introduction. Instead, we refer to the NIST ASD[1] (Kramida et al. 2018) for the best available data in the literature relating to the levels/lines in the current work. The ASD lists three experimental sources for many of the lines in this study: Bergeson et al. (1996, hereafter B+96), Schnabel et al. (2004, hereafter SSK04), and unpublished material from Ward Whaling utilized in the critical compilation of Fuhr and Wiese (2006). In addition, the ASD lists two theory sources for a number of lines that do not have previous experimental transition probabilities. Some of these come from the Configuration Interaction calculations of Donnelly & Hibbert (2001) and some from the Orthogonal Operator calculations of Raassen & Uylings (1998). Comparison to these published data sources will be made below.

We concentrate, in this study, on measuring BFs for mostly UV lines originating in the low-lying odd-parity levels of Fe II. These levels range in energy from 38459 – 47626 $cm^{-1}$ and belong to the $z^6D^o$, $z^6F^o$, $z^6P^o$, $z^4F^o$, $z^4D^o$ and $z^4P^o$ terms of the $3d^6(^5D)4p$ configuration. These UV lines represent >98% of the radiative decay from each of these levels. The current study will set the foundation for the further measurement of weak and very weak optical branches from these same upper levels. Our BF measurements are summarized in Section 2 below. These BFs are converted to transition probabilities using radiative lifetimes from the literature in Section 3. Comparison between the published sources in the NIST ASD and our results are made in Section

[1] Atomic Spectra Database of the National Institute of Standards and Technology

4 of this paper.  Finally in section 5, we apply our new data to the determination of abundances in the Sun and in metal-poor star HD 84937.

## 2. BRANCHING FRACTION MEASUREMENTS IN Fe II

The BF for a transition between an upper level $u$ and lower level $l$ is the ratio of its A-value to the sum of all the A-values associated with $u$.  This can also be expressed as the ratio of relative emission intensities I for these transitions:

$$BF_{ul} = \frac{A_{ul}}{\sum_l A_{ul}} = \frac{I_{ul}}{\sum_l I_{ul}}. \quad (1)$$

The BFs reported here are determined in large part from several low current (15 – 20 mA) emission spectra produced using sealed, commercially available Fe-Ne and Fe-Ar hollow cathode discharge (HCD) lamps and recorded using the UW 3-meter echelle spectrograph.  The echelle spectrograph, operating at a resolving power of 250,000, requires three overlapping CCD frames to obtain a full spectrum in the UV.  We routinely use five frames to provide redundancy and check for lamp drift.  The sealed HCD lamps run very stably and reproducibly at low currents over the many hours required to obtain a complete spectrum.  Operating at low current is essential in order to avoid optical depth on the strongest lines.  The spectra taken with the 3-meter echelle are listed in Table 1.  These spectra are all calibrated using a NIST traceable deuterium ($D_2$) standard lamp.  A spectrum from this $D_2$ lamp is recorded immediately after each HCD spectrum.  This "everyday" $D_2$ lamp is also periodically checked against a little-used NIST traceable $D_2$ lamp to correct for lamp aging.  The calibration of the $D_2$ lamp contributes <2% to the uncertainty of the relative line intensities.

One shortcoming of using sealed commercial HCD lamps as an emission source, particularly in the UV, is that it is difficult to correct the spectra for changes in window transmission as a function of wavelength without sacrificing the lamp.  In the current study, the echelle spectra are utilized over a fairly narrow wavelength range, 2250 – 3280 Å, but there is still likely some falloff in transmission over this range due to a combination of the quality of the surface polish and the potential buildup of a sputtered metal thin film on the inner surface.  In order to correct the wider set of data taken with the sealed HCD lamps, a partial Fe-Ne UV spectrum was measured on a single CCD frame using a custom built see-through HCD running at 40 mA.  This is the final spectrum listed in Table 1.  The window transmission of this custom HCD was measured by comparing a $D_2$ lamp spectrum with and without the see-through HCD in the light path. This works because the lamp has spatial symmetry and we can assume any metal film that has grown on the front window will be identical to that on the back window.  This single frame spectrum, corrected for window transmission, is used to determine a limited set of branching ratios (BRs) from the $z^4D^o_{5/2}$, $z^4D^o_{3/2}$, $z^4F^o_{7/2}$ and $z^4F^o_{5/2}$ levels which all have branches within ~10 Å of 2250 Å, 2370 Å, 2740 Å and 3200 Å.  These BRs, in turn, are used to make corrections to the farthest UV data from the sealed HCD spectra.  These corrections amount to ~12% at 2250 Å and ~3% at 2370 Å.  No correction is needed for wavelengths longer than 2430 Å.  The UV and blue lines in this study represent almost all of the radiation from these upper levels, but an estimate of residuals was made from the Kurucz & Bell (1995) database[2] and corrections of less than 2% were made accordingly.

As stated above, the branching fractions for most of the lines in this study were determined entirely from the echelle spectra listed in Table 1. Exceptions to this are the lines connecting to the $z^4P$ levels and the blue lines associated with the $z^4D$ levels. The levels of the $z^4P$ term were studied using a combination of the echelle spectra in Table 1 and the first 10 FTS spectra listed in Table 2. The FTS spectra were all taken on the (now decommissioned) 1 meter FTS on the McMath-Pierce Solar Telescope at National Solar Observatory, Kitt Peak and are publically available for download.[3] Ar II branching ratios from Whaling et al. (1993) were used to determine the slope of the FTS instrument response across the $z^4P^o$ – $a^4P$ multiplet at ~2970 Å. The wavelength spread for lines in this multiplet is small, ~58 Å, so the change in instrument response is minimal within the multiplet. The Ar II branching ratios do not extend far enough into the UV to do the same for the far-UV multiplet at ~2580 Å. Instead, the slope of the instrument sensitivity was minimally adjusted over the ~49 Å spread to bring the multiplet into best agreement with LS theory. LS theory should be quite good for this multiplet as the $z^4P^o$ upper levels are 96% pure and the $a^4D$ lower levels are 99% pure. Data from the echelle spectra in Table 1 was used to calibrate the FTS data so that lines of the deep UV multiplet and near UV multiplet could be put on a common intensity scale.

In the case of the $z^4D$ levels, the far UV and near UV branching fractions were determined from the low current echelle spectra. The strength of the weak blue lines relative to that of the near-UV lines was measured using the final 17 spectra listed in Table 2. These spectra range in current from 420 – 1300 mA. The relatively high current is necessary in order to bring up the weak blue lines of the $z^4D^o$ – $b^4P$ multiplet with adequate signal-to-noise, but the high current results in the strong branches connecting to the lower $a^4D$ and $a^4F$ terms being optically thick in these spectra. Great care was taken to make sure that the near-UV lines at ~3200 Å, which are much weaker (1 – 2%) branches connecting to the higher-lying $a^4P$ term, are optically thin in order to produce reliable branching ratios with the weak blue lines connecting to the $b^4P$ term.

The uncertainties on final mean BFs are evaluated from the strength of the BFs, the S/N of the spectral lines, and the wavenumber difference of lines from the common upper level. By definition, branching fractions from a given upper level sum to 1.00, so uncertainty migrates to the weaker lines. The relative radiometric calibration also contributes to the uncertainty. The conservative estimate of the calibration uncertainty is 0.001%/cm$^{-1}$ of the wavenumber difference between a line and the dominant branch(es) from the common upper level (Wickliffe et al. 2000).

## 2.1 Blend Separation

As mentioned above, the emission spectra we use for the branching fraction work are produced in HCDs operating with either Ar or Ne buffer gas. In addition to the desired lines of Fe II, the spectra will contain lines of Fe I as well as lines of Ar I & II or Ne I & II. There are a number of instances where the wavelength of an Fe II line of interest is a near or exact match with the

[2] available online at https://www.cfa.harvard.edu/amp/ampdata/kurucz23/sekur.html
[3] FTS data are publicly available at http://diglib.nso.edu/

wavelength of another Fe II line, a line of Fe I or a buffer gas line. If the Fe II line is blended with an Ar I or II line in the Fe-Ar HCD spectra, one simply determines its branching fraction solely from the Fe-Ne HCD spectra, and vice versa. In the case of a potential blend with Fe I or Fe II, we first looked for evidence to determine if it is a real blend. The NIST ASD was consulted to see what specie(s) and energy levels the line is associated with in their line list. The most telling evidence of a blend is to look at whether its branching ratio changes with different lamp conditions. Blends between ions and neutrals are commonly studied by measuring the ratio of the blended line to a clean line from the same upper level over a range of lamp currents. This works because ion and neutral line strengths exhibit a different current dependence. This technique requires that both lines involved in the blend remain optically thin over the range of currents studied. This technique is not useful for studying ion-ion blends, as the two lines will have roughly the same rate of change with current.

In this study we have used two different techniques to separate blends. The transition at 3002.64 Å from the $z^4P^o_{5/2}$ level is separated using a center-of-gravity technique. This technique is possible because the Fe II level energies involved are known to very high accuracy (Nave & Johansson 2013; Kramida et al. 2018) and this level is one of those that we studied using a combination of echelle and FTS spectra. The internal wavenumber accuracy and precision of FTS data is better than 1 part in $10^7$. This level of accuracy is required for center-of-gravity separation. The technique simply involves comparing the center-of-gravity wavenumber of the blended feature with the Ritz wavenumbers of the blending partners. This technique works best for S/N>10 and a Ritz wavenumber separation >0.05 $cm^{-1}$. In the case of the line at 3002.64 Å the Ritz separation is 0.026 $cm^{-1}$. The line is separable, but with somewhat increased uncertainty.

The other technique used here to separate blends takes advantage of the observation that line strengths of both ions and neutrals are significantly enhanced in neon buffer gas relative to argon buffer gas, all other discharge parameters being equal. Not only are lines enhanced in neon, but the factor by which they are enhanced varies widely as a function of species and upper level. As a result the two lines that make up a blend will have different blending fractions in neon and argon. By comparing ratios of the intensity of the blend to the intensities of clean lines from the upper level of each blending partner, one can solve for the blending fraction in each spectra. This is an exact solution if comparing one pair of ratios for each spectrum, or one can use a least-squares optimization when comparing multiple blend to clean line ratios for each spectrum. The strength of this technique is that it can be successful with as few as two spectra, one with neon and one with argon buffer. It does not require a whole range of currents, which might lead to optical depth in some lines, but can be carried out on low current spectra that are sure to be optically thin. Occasionally the upper level of a blending partner has no other clean transition with which to compare the blend. We have observed that, with very few exceptions, the levels in a given term have very similar neon/argon enhancement. In this case it can often prove useful to compare the blend to clean lines from other upper levels from the same term. This was successfully done for the blending partner for the line at 2359.11 Å from the $z^6P^o_{5/2}$ upper level and was also used to confirm that the line at 2746.98 Å from the $z^4D^o_{5/2}$ upper level was, in fact, unblended.

The eight potential blends we have investigated are summarized in Table 3. This table lists the potential blend, the upper levels of the lines involved in the blend, as well as the clean, unblended lines that were used to compare to in the case of a least-squares optimization. The final row of Table 3 lists the only line that was separated by the center-of-gravity technique in this study. Table 3 contains extensive notes giving the details of each blend and how it was handled.

## 3. TRANSITION PROBABILITIES AND log(*gf*)s

Our measured BFs for 131 lines of the $z^6D^o$, $z^6F^o$, $z^6P^o$, $z^4F^o$, $z^4D^o$ and $z^4P^o$ levels are reported in Table 4 organized by term-ordered upper levels. Our BFs are converted to transition probabilities and subsequently to log(*gf*)s using the relations from Thorne (1988), where $A_{ul}$ is in $s^{-1}$ and λ is in nm:

$$A_{ul} = \frac{BF_{ul}}{\tau_u} \; ; \; \log(gf) = \log(1.499 \times 10^{-14} \lambda^2 g_u A_{ul}) \qquad (2)$$

The lifetimes for most of the levels in this study have multiple good measurements reported in the literature, and we have not remeasured them. We rely most heavily on the recent measurements of SSK04, as well as Guo et al. (1992) and Biémont et al. (1991). Guo et al. and Biémont et al. used the laser – fast beam method that eliminates any need for fast electronics. SSK04 used time-resolved nonlinear laser-induced fluorescence with a stimulated Brillouin scattering technique to shorten their laser pulse. All three groups made serious attempts to control any possible systematic errors. For most levels we used a straight average of the lifetimes presented in two or all three of these works. We did not include the older lifetimes of Schade et al. (1988) in our averages as their lifetimes have significantly larger quoted uncertainties. For the $z^4D^o_{1/2}$ and $z^4F^o_{3/2}$ levels we use the only available lifetimes from Hannaford et al. (1992) and for the $z^4P^o$ levels we use a lifetime of 3.25 ns for all three levels. This value is closest to the SSK04 measurement for the J=5/2 and 3/2 levels, but has overlapping uncertainties with Guo et al. and Li et al. (1999) for those levels. The J=1/2 level does not have a lifetime measurement available in the literature. Table 5 summarizes the radiative lifetime values used and the source(s) of those values. The radiative lifetimes in Table 5 are combined with our BFs in Table 4 to determine A-values and log(*gf*)s for 131 transitions of Fe II. These are presented in Table 6, which is available in machine readable form.

## 4. COMPARISON WITH PREVIOUS EXPERIMENT AND THEORY

Many of these lines have been investigated experimentally in the past. The NIST ASD lists the studies by B+96 and SSK04 as the primary sources of experimental transition probabilities for many of these lines, although five of the levels in the current study were omitted from the B+96 study, and three from the SSK04 study. Two levels, the $z^6P^o_{5/2}$ and $z^4P^o_{1/2}$, were omitted from both earlier experimental studies and only theoretical results are available in the NIST ASD for lines from these levels. We include results from the B+96 and SSK04 studies in Table 4. Because we are reporting new branching fractions and are combining these with radiative lifetimes from the literature, we make comparison to the BFs from these earlier studies rather than to the final transition probabilities.

The B+96 work resulted from a collaboration between our (UW) group and the group at University of Lund. Branching fractions were measured using a combination of FTS spectra and grating spectrometer spectra, much as for the $z^4P^o$ levels in the current study. The present work is more comprehensive than the earlier B+96 study, with more levels included and additional weak lines for many of the levels. B+96 did not report branching fractions for the $z^6P^o_{5/2}$ and $z^4D^o_{5/2}$ levels because of blends on major branches that they found intractable (see discussion of blends in 2.1 above and in Table 3), although they did report branching ratios for some other branches from these levels. They also did not study the $z^4P^o$ levels. Figure 1 shows a comparison for the 67 lines that the studies have in common, with $\log(BF)_{B+96} - \log(BF)_{this\ exp.}$ plotted versus $\log(BF)_{this\ exp.}$ in panel (a) and versus wavelength in panel (b). The error bars represent the uncertainties of both studies combined in quadrature. The solid line at zero represents perfect agreement. As can be seen from these plots, the agreement between these two studies is excellent with all lines lying within 0.1 dex, and most much better.

The SSK04 study reported new radiative lifetimes for these levels but relied on older BFs from earlier publications from their group: the Ph.D. Thesis of Kroll (1985), and the journal papers by Kroll & Kock (1987) and Heise & Kock (1990). The Kroll & Kock study reports on the UV transition probabilities while the Heise & Kock publication deals with the very weak optical lines from these levels. The emission BFs reported by Kroll & Kock were made with a combination of FTS and grating spectrometer data. However, the FTS spectra they used exhibited substantial optical depth for the strong lines. They applied corrections to the integrated intensities of these lines by comparing the line width of the optically thick line to the Doppler width measured from an optically thin line. Such corrections are difficult and cause substantial added uncertainty. Figure 2 shows a comparison of our BFs to those reported in SSK04 that overlap our study. As with Figure 1, the error bars represent the uncertainties of both studies combined in quadrature. The solid line at zero represents perfect agreement. The plots in Figure 2 are put on the same vertical scale as Figure 1 to aid the eye (although the horizontal scale is expanded). The comparison here is less favorable than the comparison to B+96, although most lines do lie within the (larger) combined uncertainties of the data.

There are a number of lines in our study that are not reported by either B+96 or SSK04. The NIST ASD sources transition probabilities for five of these lines from the Configuration Interaction calculations of Donnelly & Hibbert (2001, hereafter DH01) and 17 from the Orthogonal Operator calculations of Raassen & Uylings (1998, hereafter RU98). The lines that are sourced from DH01 in the NIST ASD are the $z^6F^o_{5/2} - a^6D_{7/2}$ transition at 2383.06 Å, the $z^6P^o_{5/2} - a^6D_{7/2,5/2,3/2}$ transitions at 2332.80, 2348.30 and 2359.11 Å and the $z^4F^o_{3/2} - a^6D_{3/2}$ transition at 2250.18 Å, all of which are given a C+ rating (somewhat better than ±25%) in the ASD. The remaining 17 transitions that are sourced from RU98 in the ASD are mostly weak, intercombination lines (except for the four lines from the $z^4P^o_{1/2}$ level) and all have a NIST rating ranging between D+ and E (D+: somewhat better than ±50%; E: greater than 50% but less than factor of 3). Figure 3 shows a comparison of these calculated values with our log(*gf*)s. As in Figures 1 and 2, the solid horizontal line represents perfect agreement.

## 5. APPLICATION TO SOLAR AND STELLAR IRON ABUNDANCES

We apply the Fe II laboratory transition probabilities to derive new abundances in the solar photosphere and in the metal-poor main sequence turnoff halo star HD 84937. In all aspects our analysis follows our previous iron-group species studies (Sneden et al. 2016 and references therein, Lawler et al. 2017, 2018, 2019).

### 5.1 Fe II in the Solar Photosphere

A solar abundance re-determination has been featured in our past papers, but it can only play a supporting role here, due to lack of enough suitable Fe II lines. All transitions in Table 6 with wavelengths $\lambda < 4000$ Å suffer from some combination of: (a) being so strong in the Sun that they lie well up on the flat and damping parts of the c-o-g; (b) existing in the extremely crowded *UV* spectral region, with attendant problems of line blending and uncertain continuum levels; and (c) lacking a suitable solar high-resolution spectral atlas below 3000 Å. Therefore we are left with 10 potentially useful lines in the blue spectral region, 4173 Å $\leq \lambda \leq$ 4620 Å. These lines are also strong. They are listed in the Moore et al. (1966) solar line compendium with equivalent widths *EW* ranging from 47–231 mÅ. After deblending of the Fe II lines from contaminants, Moore et al. derived reduced widths log(*RW*) = log(EW/$\lambda$) from –5.00 to –4.37, essentially putting all lines on the flat part of the c-o-g. In short, all of these transitions are at least somewhat saturated, and thus are sensitive to microturbulent velocity choice and damping parameters. Fortunately, damping is not a serious issue here, as all of the Fe II transitions in Table 6 were included in the comprehensive damping calculations of Barklem & Aspelund-Johansson (2005); their results are adopted here.

Briefly summarizing our solar photospheric abundance procedure here, we use synthetic spectra to match the observed center-of-disk solar spectrum of Delbouille et al. (1973).[4] The synthetic spectrum line lists are generated with the atomic lab data of the Wisconsin group and molecular data of the Bernath group (e.g. Masseron et al. 2014, Sneden et al. 2014), supplemented by transitions in the Kurucz (2011,2018) compendium[5]. Our transition choices are discussed in previous papers of this series, and are available in an on-line resource curated by V. Placco.[6] We adopt the solar photospheric model of Holweger & Müller (1974) in order to maintain consistency with our previous analyses. This model atmosphere and line lists are inputs to the local thermodynamic equilibrium (LTE) plane-parallel spectum line analysis code MOOG (Sneden 1973)[7] to produce synthetic spectra, and these are broadened empirically with Gaussian functions to compare with the observed spectra.

We compare initial synthetic and observed photospheric spectra for all 21 Fe II lines with $\lambda \geq$ 3170 Å in Table 6. These tests immediately result in elimination of the 11 near-*UV* lines (3170–3227 Å). Each of these lines is very strong and usually severely blended with other transitions, making it impossible to determine a reliable solar Fe abundance. Among the 10 longer-wavelength lines, we also eliminate the lines at 4303.16 and 4351.76 Å; they are compromised by CH and other atomic contaminants. For the remaining 8 blue Fe II lines we list the individual abundances in Table 7. Adopting standard spectroscopic abundance definitions[8], we derive <log $\varepsilon(Fe)_\odot$> = 7.46 ± 0.03 (σ = 0.08). This abundance is in good agreement with current recommended Fe photospheric abundance, log ε = 7.50 ± 0.04 (Asplund et al. 2009), 7.51 ± 0.04 (Scott et al. 2015), and with the meteoritic abundance, 7.45 ± 0.01 (Lodders et al. 2009).

[4] http://bass2000.obspm.fr/solar_spect.php
[5] http://kurucz.harvard.edu/linelists.html
[6] https://github.com/vmplacco/linemake
[7] http://www.as.utexas.edu/~chris/moog.html

Given the lack of comprehensive lab studies of Fe II transitions in the past couple of decades, many abundance studies have chosen to adopt the transition probabilities recommended either by (a) NIST[9] from the critical compilation of Fuhr & Wiese (2006), or by (b) Meléndez & Barbuy (2009; hereafter MB09), constructed from a variety of lab and theoretical sources and from fitting the solar spectrum; we will call them "empirical" log(*gf*) values. For the eight solar blue Fe II lines in common with our study, the NIST log(*gf*) values are about 0.1 dex larger than those in Table 6, and with them we derive <log $\varepsilon(Fe)_\odot$> = 7.31 (σ = 0.11). The MB09 log(*gf*)'s for these lines are similar to ours, from which we calculate <log $\varepsilon(Fe)_\odot$> = 7.52 (σ = 0.08). For this very restricted set of Fe II transitions we agree well with MB09.

### 5.2 Fe II in HD 84937

For HD 84937 a much more extended analysis is possible. This star has well-determined atmospheric parameters: effective temperature $T_{eff}$ = 6300 K, surface gravity log $g$ = 4.0, metallicity [Fe/H] = –2.15, and microturbulent velocity $v_t$ = 1.5 km $s^{-1}$. It has been studied extensively with high-resolution spectroscopy; in the PASTEL stellar spectroscopic bibliography (Soubiran et al. 2016) HD 84937 has more than 40 entries[10]. The adopted atmospheric parameters have been discussed in detail by Sneden et al. (2016) and will not be repeated here; they are in good accord with the previous investigations listed in the PASTEL database. A recent reexamination of the HD 84937 parameters by Roederer et al. (2018) yields $T_{eff}$ = 6418 ± 117 K, log $g$ = 4.16 ± 0.14, [Fe/H] = –2.25 ± 0.1, and $v_t$ = 1.5 ± 0.2 km $s^{-1}$, we keep the original Sneden et al. (2016) parameters here to be consistent with previous papers of this series.

The observed HD 84937 spectra are the same ones that we have employed in past papers of this series. In the vacuum-*UV* (2300 Å ≤ λ ≤ 3100 Å) we use the archival Hubble Space Telescope Imaging Spectrograph (HST/STIS) obtained under proposal GO7402 (PI: R. C. Peterson), which has R ≡ λ/Δλ ≈ 25,000, and typical signal-to-noise S/N ~ 50. In the near-UV (defined here for convenience as 3100 Å ≤ λ ≤ 3300 Å), and in the optical (3300 Å ≤ λ ≤ 6500 Å) we use an archival ground-based ESO Very Large Telescope UVES spectrum with R ≈ 60,000 and typical S/N ~ 100 in the blue and ~300 in the red. Synthetic spectra are computed with atomic/molecular line lists generated in the manner described in §5.1, with a model atmosphere interpolated from the ATLAS grid (Kurucz 2011, 2018)[11].

---

[8] For elements X and Y, the relative abundances are written [X/Y] = $\log_{10}(N_X/N_Y)_{star} - \log_{10}(N_X/N_Y)_\odot$. For element X, the "absolute" abundance is written log ε(X) = $\log_{10}(N_X/N_H) + 12$. Metallicity is defined as [Fe/H], a logarithmic ratio of element densities normalized to 0 for the Sun.

[9] https://physics.nist.gov/PhysRefData/ASD/lines_form.html

[10] http://vizier.u-strasbg.fr/viz-bin/VizieR?-source=B/pastel

[11] http://kurucz.harvard.edu/grids.html

Given the excellent HST/STIS spectrum of HD 84937, all of the transitions in Table 6 with $\lambda >$ 2300 Å are potential Fe abundance candidates. We form synthetic spectrum line lists in the manner described in §5.1. From the synthetic/observed matches we identify 66 useful lines with $\lambda \leq 3300$ Å. None of these features are weak, and they all lie on the flat or damping portions of the HD 84937 c-o-g. However, they cover an enormous strength range. We illustrate this with three examples in Figure 4.

In panel (a) of Figure 4 we show the lowest-wavelength line, 2327.40 Å. The red synthetic spectrum is computed with no contribution from Fe I or Fe II lines, and it illustrates that the line of interest here is essentially uncontaminated with transitions of other species. The black line illustrates the best-estimate Fe abundance (Table 7) for this line. The blue, green, and orange lines show syntheses for different assumed Fe abundances, offset from this best estimate by –0.6, –0.3, and +0.3, respectively. The resulting predicted line strengths are not very different than the best fit, because this transition is saturated (log(*RW*) ~ –4.0), well along the "flat" c-o-g region for HD 84937, where line equivalent widths grow very slowly, approximately as $(\ln N_{Fe})^{1/2}$. In panel (b) of Figure 4 we show the strongest Fe II line, 2599.40 Å (log(*RW*) ~ –3.5), and its significant blending companion 2598.37 Å (log(*RW*) ~ –3.7). Comparison of the synthetic spectra of these lines with that of 2327 Å in panel (a) shows that they have greater responsiveness to abundance changes, because lines in the c-o-g damping regime grow approximately as $N_{Fe}^{1/2}$. Finally, in panel (c) we feature 2917.46 Å, one of the weakest Fe II vacuum-UV transitions that we are able to use. This line (log(*RW*) ~ –4.6) is obviously much weaker than the ones displayed in Figure 4(a,b), but by inspection it is clearly blended. The contaminant is V II 2917.36 Å, whose laboratory transition parameters are well established (Wood et al. 2014a). We adopt the V abundance from that study for the synthesis shown here.

Neither the *UV* Fe II lines of Figure 4 nor most others not discussed individually match the ideal characteristics for abundance analysis – none of them are intrinsically weak, and many of them have minor or major blending issues in this crowded spectral region. Therefore we cannot be too selective here. Our goal is to test for line-to-line consistency in derived Fe abundances, and so we retain all lines that are not severely compromised by blends from Fe I and other species transitions. From 66 vacuum-UV lines meeting these criteria (Table 6) we derive $\langle \log \varepsilon(\mathrm{Fe}) \rangle =$ 5.26 ($\sigma = 0.07$) for HD 84937. We plot these abundances with wavelength in the top panel of Figure 5, defining the UV spectral range as $\lambda < 3300$ Å.

In the "blue" domain (4170 Å $\leq \lambda \leq$ 4625 Å) of the optical region, the 10 Table 6 lines are all weak and mostly unblended. The 4351.76 Å line must be discarded as it suffers from multiple blends. The 4549.47 Å line has significant blending from Ti II 4549.64 Å, but an Fe abundance can be derived from our synthetic spectra in HD 84937. All of the lines have log(*RW*) ≤ –5.0, nearly all of them qualifying for membership in the linear portion of the c-o-g for this star. As in past papers of this series a relative line strength indicator can be defined as $\log(gf) - \theta\chi$, where $\theta = 5040/T$ and $\chi$ is the lower excitation energy of the line. For the 66 *UV* lines, approximating $\theta \approx 0.8$, $\langle$strength$\rangle \approx -1.7$, while for the 9 blue lines $\langle$strength$\rangle \approx -4.5$, nearly three orders of magnitude weaker than the *UV* lines. Abundance derivation from the blue lines from synthetic/observed spectrum comparisons is straightforward (Table 7), and we determine $\langle \log \varepsilon(\mathrm{Fe}) \rangle = 5.28$ ($\sigma = 0.10$) for HD 84937.

MB09 reported empirical transition probabilities for many lines that we can detect on our *VLT/UVES* spectrum of HD 84937. For blue lines in common with our study we derive <log ε(Fe)> = 5.32 (σ = 0.03) using their log(*gf*) values, in good accord with the abundance derived with our *gf*'s. Additionally, from the much larger set of 39 optical-region lines in their study that we can detect, we get <log ε(Fe)> = 5.27. These individual line abundances are plotted in panel (b) of Figure 5. The line-to-line scatter is very small, suggesting that their transition probabilities yield reliable abundances for lines in HD 84937.

Repeating the HD 84937 abundance computations with Fe II transition probabilities from NIST produces more discordant results. In the UV region, our log(*gf*) values are close to the NIST ones, leading to identical mean abundances. But in panel (c) of Figure 5 six discordant lines near 2900 Å (Table 7) are seen. They are a distinct minority of the 58 measured NIST lines in the UV, and thus do not significantly perturb the mean abundance. More problematic are the optical lines. For just the blue lines in common with our work the NIST transition probabilities yield <log ε(Fe)> = 5.15, about 0.15 dex lower than the mean abundance derived from our and MB09 log(*gf*) values. Expanding the line set to all optical lines brings the mean NIST-based abundance to <log ε(Fe)> = 5.17, slightly closer to our value, but with a large sample standard deviation, σ = 0.15. The large line-to-line scatter is apparent in panel (c) of Figure 5.

We conclude from solar and HD 84937 analyses that the MB09 transition probabilities are recommended for future Fe II abundance studies, absent new laboratory investigations. We caution the reader that our summary statement applies only to those transitions with relatively low excitation energies. Lines arising from levels with $\chi > 4$ eV have not been explored here, as they are undetectable on our HD 84937 spectra.

We had studied the complete iron-group elemental abundances in HD 84937 (Sneden et al. 2016). We argued that standard Saha ionization equilibrium constraints are well satisfied for the seven elements with detectable lines of both neutral and ionized species (Ti through Ni). We found that Sc, Ti, and V (the three lightest iron-group elements) are significantly overabundant in comparison to the heavier elements. For example, [V/Fe] = 0.25 with [Fe/H] = –2.32 derived from both species in that paper. For Fe II lines, Sneden et al. adopted NIST log(*gf*) values. Here we have determined log ε = 5.26, or [Fe/H] = –2.24[12] with the new lab transition probabilities. This 0.08 upward shift in Fe II abundances would shift all [X/Fe] ionized-based abundances downward by that amount. The small discrepancy between Fe I and Fe II abundances in HD 84937 could be easily erased via a decrease in surface gravity by ~0.15 dex, a very small shift that is within the uncertainty of the adopted model atmosphere gravity.

The newly derived Fe II determinations will not affect the overall correlations among the iron-group elements, but will mute the overabundances of Sc, Ti, and V slightly. No changes to the abundances from neutral species will result from our present Fe II study. Finally, these new Fe II abundances will be important for all such Galactic Chemical Evolution studies, enabling determination of accurate abundance ratios in main sequence turnoff stars that can directly constrain nucleosynthetic predictions in massive stars (Cowan et al. 2019).

[12] For the solar Fe abundance we use the commonly adopted log ε = 7.50 rather than 7.46 that we derived in §5.1, because our log ε estimate in HD 84937 is dominated by UV transitions that are unavailable in the Sun.

## 6. SUMMARY

We report new branching fractions for 121 UV lines from the low-lying odd-parity levels of Fe II belonging to the $z^6D^o$, $z^6F^o$, $z^6P^o$, $z^4F^o$, $z^4D^o$ and $z^4P^o$ terms of the $3d^6(^5D)4p$ configuration. These lines range in wavelength from 2250 – 3280 Å and originate in levels ranging in energy from 38459 – 47626 $cm^{-1}$. In addition, we report branching fractions for 10 weak blue lines originating from the $z^4D$ levels. These BFs are combined with radiative lifetimes from the literature to determine transition probabilities and log(*gf*) values. Our new data are compared to previous experimental and theoretical data which appear currently in the NIST ASD (Kramida et al. 2018). Our new log(*gf*)s are applied to the determination of the iron abundances in the Sun and in the metal-poor star HD 84937.

## ACKNOWLEDGEMENTS

This work is supported in part by NASA grant NNX16AE96G (J.E.L.), by NSF grants AST-1516182 and AST-1814512 (J.E.L. & E.D.H.), and NSF grant AST-1616040 (C.S.).

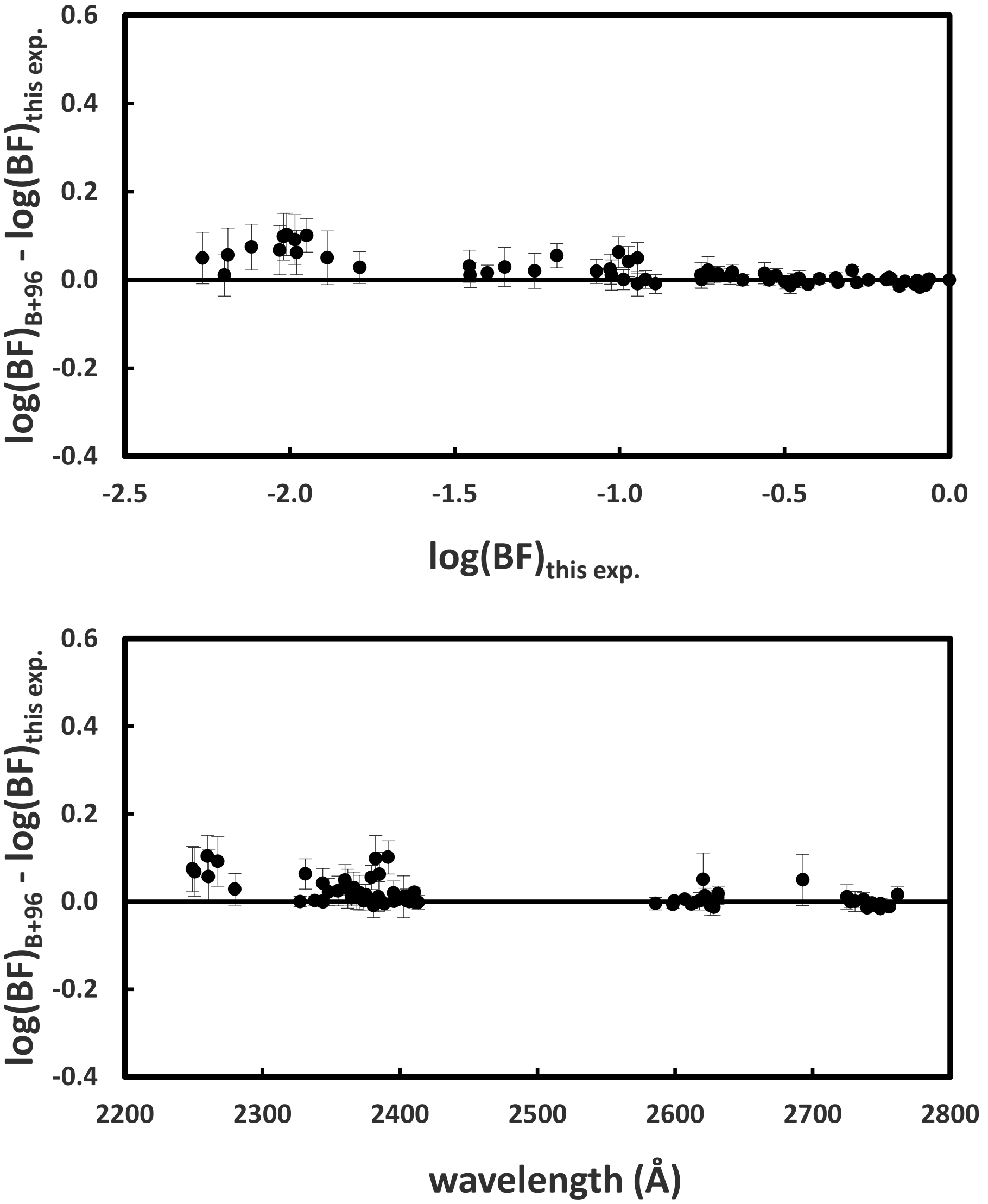

Figure 1. (a) Comparison of log(BF) for 67 lines from B+96 with those from this experiment as a function of $\log(\mathrm{BF})_{\text{this exp.}}$. The solid line at zero indicates perfect agreement. (b) same as for (a) except versus wavelength.

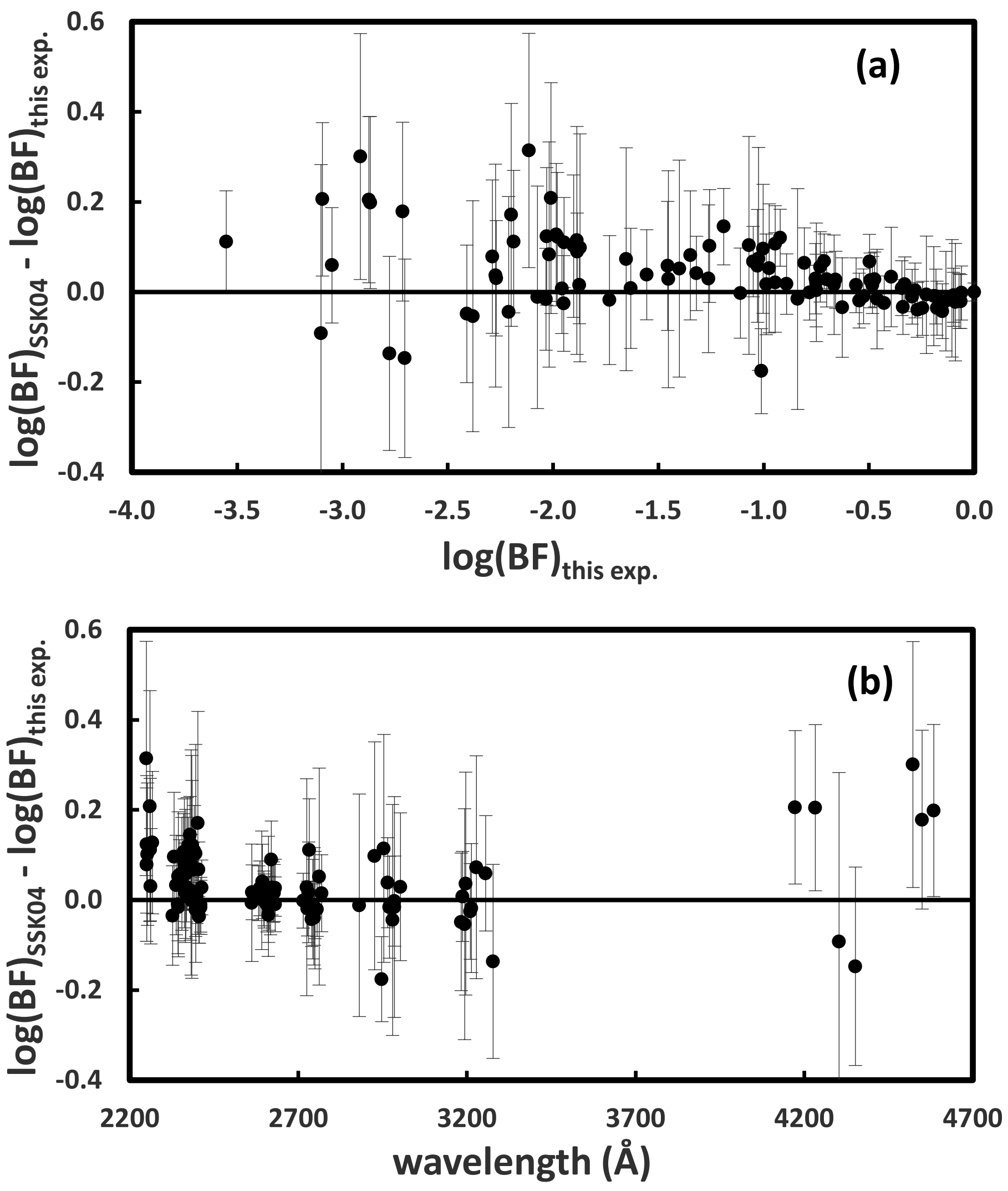

Figure 2. (a) Comparison of log(BF) for 96 lines from SSK04 with those from this experiment as a function of $\log(BF)_{\text{this exp.}}$. The solid line at zero indicates perfect agreement. (b) same as for (a) except versus wavelength.

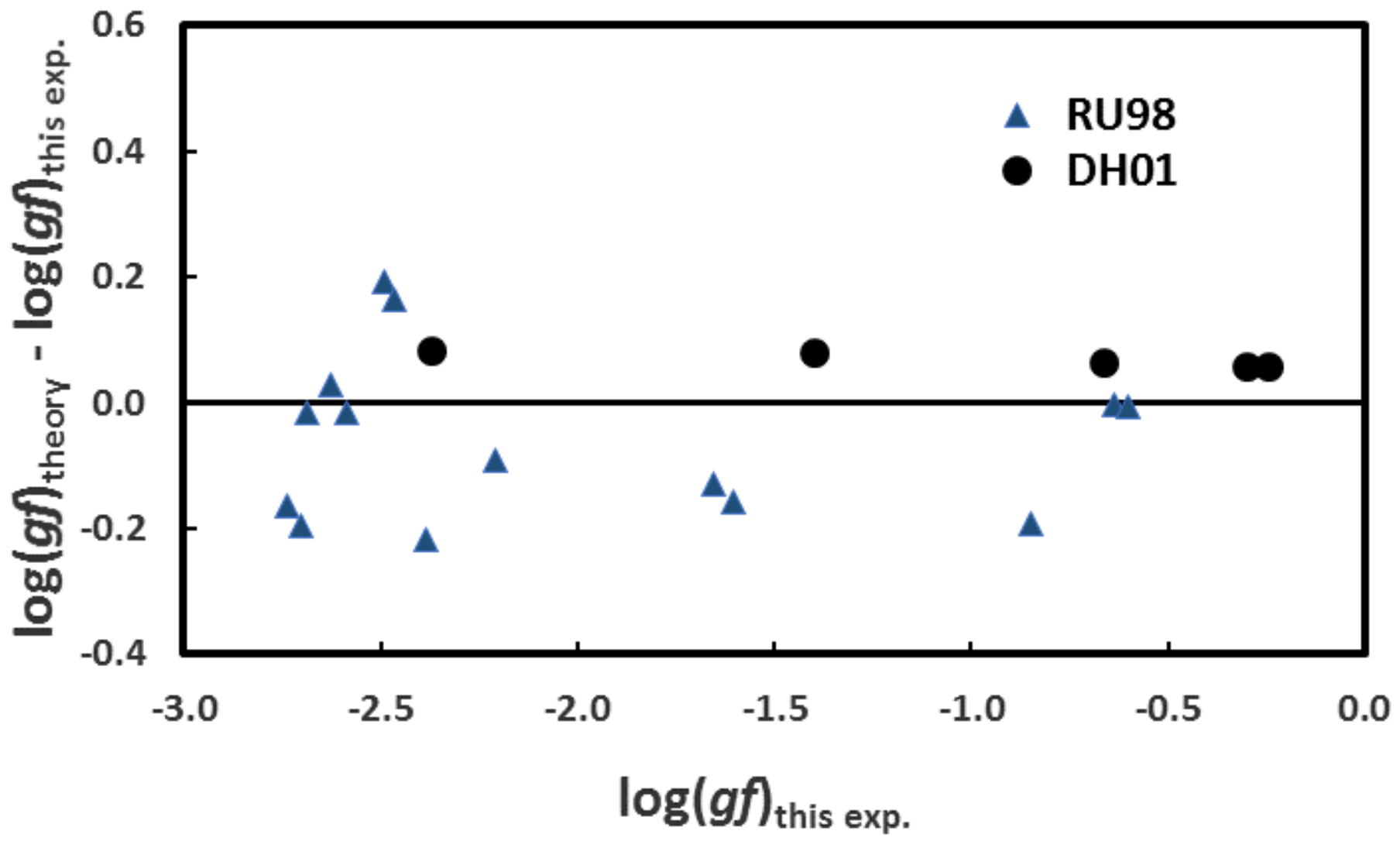

Figure 3. Comparison between our experimental log(*gf*)s and theoretical log(*gf*)s that appear in the NIST database for lines from this study. These are lines which have no previous experimental results. The blue diamonds indicate results from Raassen & Uylings (1998) and black circles from Donnelly & Hibbert (2001).

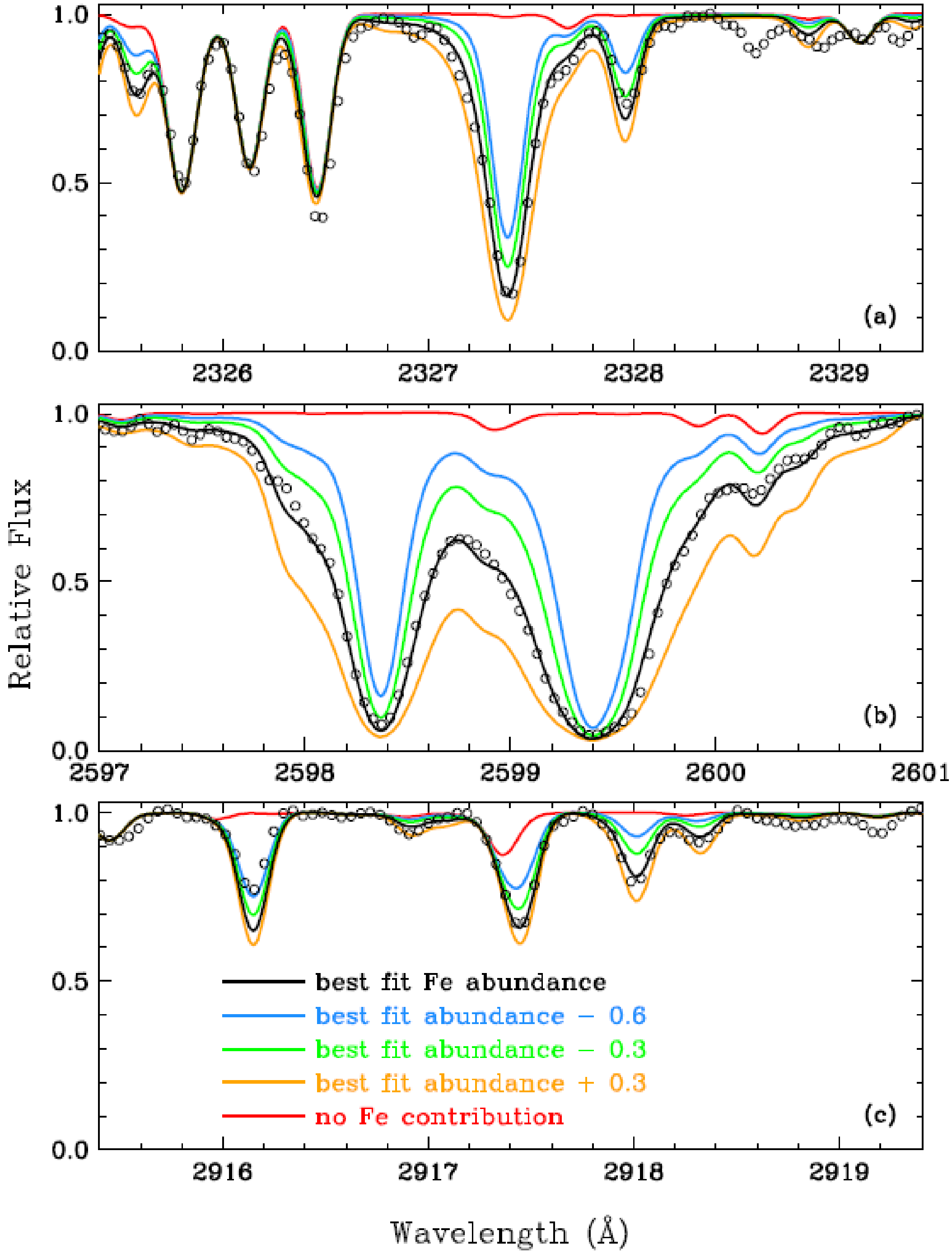

Figure 4. Observed and synthetic spectra of three Fe II transitions that have been chosen to illustrate the very large line strength range in the UV spectral region. In each panel the black open circles are the observed HST/STIS spectra. The lines represent the synthetic spectra, color coded as defined in panel (c). The blue, green, black, and orange syntheses are separated by 0.3 dex in assumed Fe abundance.

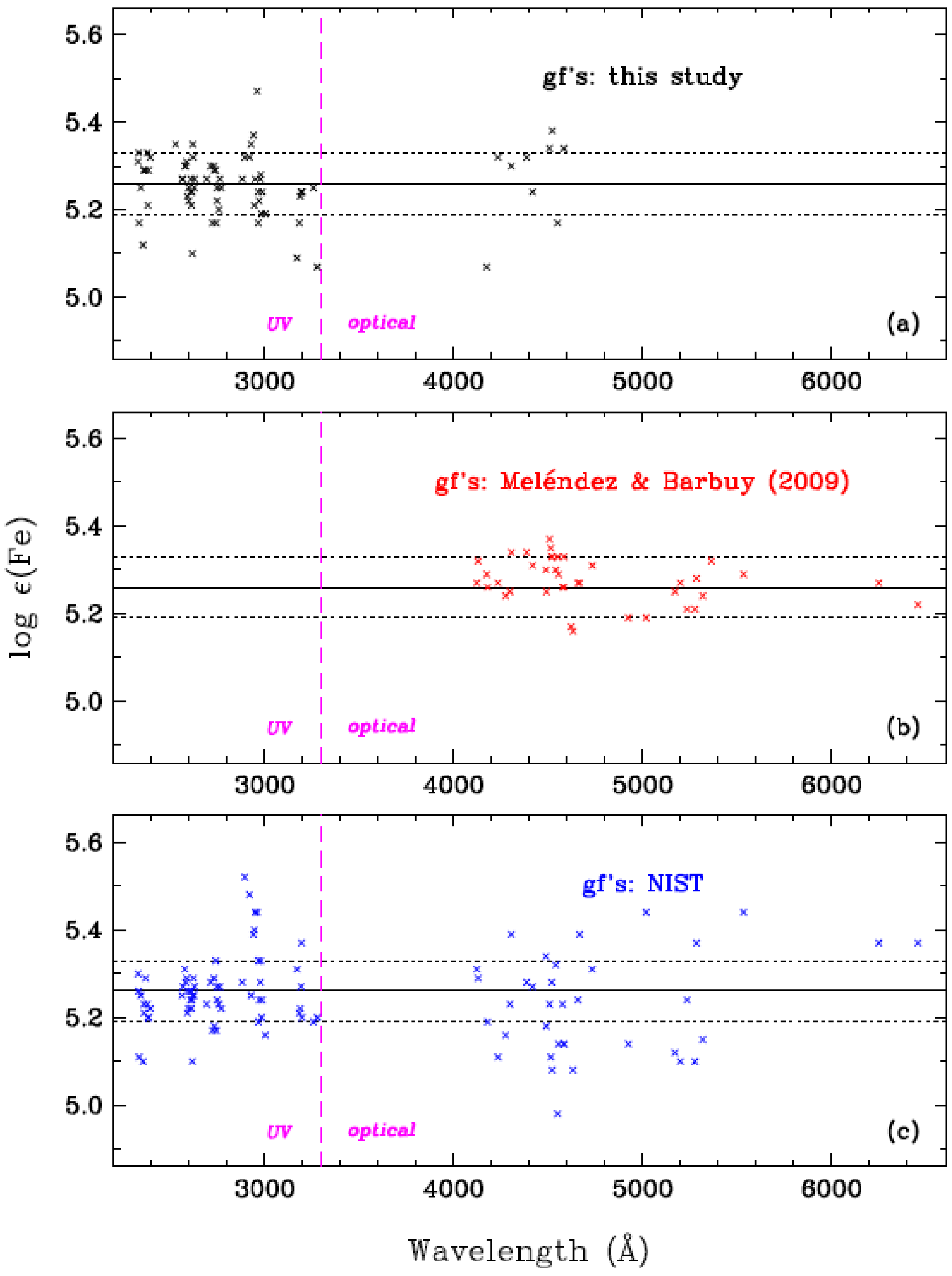

Figure 5. Fe abundances in HD 84937 derived with the transition probabilities determined in the present lab study (panel a), by the empirical computations of Meléndez & Barbuy (2009; panel b), and those recommended by NIST (panel c). The magenta lines in each panel are arbitrarily drawn at $\lambda = 3250$ Å to separate the UV and optical regions of our Fe II lines. In panel (a) the solid line represents the mean abundance and the dotted lines show the sample deviations $\sigma$ from the mean. In panels (b) and (c) we copy these same lines so that visual comparison can be made with the Meléndez & Barbuy and the NIST transition probability sets.

Table 1. Echelle spectra of commercial Fe HCD lamps and a custom see-through HCD.[a]

| Index | Date | Serial Number | Lamp Type and Buffer Gas | Lamp Current (mA) | Coadds | Total Exposure (min) |
|---|---|---|---|---|---|---|
| 58 | 2017 July 25 | 1 | commercial Ne | 15 | 48 | 20 |
| 59 | 2017 July 25 | 3 | commercial Ne | 15 | 37 | 40 |
| 60 | 2017 July 25 | 5 | commercial Ne | 15 | 96 | 40 |
| 61 | 2017 July 25 | 7 | commercial Ne | 15 | 69 | 40 |
| 62 | 2017 July 25 | 9 | commercial Ne | 15 | 96 | 40 |
| 63 | 2017 July 31 | 1 | commercial Ar | 15 | 27 | 45 |
| 64 | 2017 July 31 | 3 | commercial Ar | 15 | 10 | 100 |
| 65 | 2017 July 31 | 5 | commercial Ar | 15 | 30 | 50 |
| 66 | 2017 July 31 | 7 | commercial Ar | 15 | 24 | 60 |
| 67 | 2017 July 31 | 9 | commercial Ar | 15 | 27 | 45 |
| 68 | 2017 Aug 01 | 1 | commercial Ar | 20 | 22 | 20 |
| 69 | 2017 Aug 01 | 3 | commercial Ar | 20 | 10 | 40 |
| 70 | 2017 Aug 01 | 5 | commercial Ar | 20 | 48 | 40 |
| 71 | 2017 Aug 01 | 7 | commercial Ar | 20 | 40 | 40 |
| 72 | 2017 Aug 01 | 9 | commercial Ar | 20 | 72 | 60 |
| 73 | 2017 Aug 07 | 1 | commercial Ne | 20 | 180 | 30 |
| 74 | 2017 Aug 07 | 3 | commercial Ne | 20 | 80 | 60 |
| 75 | 2017 Aug 07 | 5 | commercial Ne | 20 | 327 | 60 |
| 76 | 2017 Aug 07 | 7 | commercial Ne | 20 | 200 | 60 |
| 77 | 2017 Aug 07 | 9 | commercial Ne | 20 | 360 | 60 |
| 78[b] | 2019 Jan 18 | 1 | custom see-through Ne | 40 | 818 | 60 |

Notes:
[a]All echelle spectra cover the wavelength range from 2200 Å to 3700 Å and have a spectral resolving power of 250,000. Three CCD frames are needed to capture a complete echelle grating order in the UV. Five CCD frames are used for each lamp operating condition to provide redundancy and a check for lamp drift. All of these spectra were calibrated with $D_2$ lamp spectra, which was recorded immediately following the completion of each HCD spectrum.
[b]The final single frame spectrum taken with a custom see-through HCD is used to determine corrections for window transmission losses in the far UV for the commercial lamp spectra, as described in the text.

Table 2
Fourier transform spectra[a] of Fe hollow cathode lamps

| Index | Date | Ser. No. | Buffer Gas | Lamp Current (mA) | Wavenumber Range ($cm^{-1}$) | Limit of Res. ($cm^{-1}$) | Coadds | Detector[b] |
|---|---|---|---|---|---|---|---|---|
| spectra for the study of the UV multiplets from the $z^4P^o$ term[c] | | | | | | | | |
| 1 | 1984 Dec. 8 | 1 | Ar | 800 | 30397 - 45035 | 0.029 | 8 | Solar Blind PMT |
| 2 | 1984 Dec. 8 | 2 | Ar | 100 | 30397 - 45035 | 0.029 | 8 | Solar Blind PMT |
| 3 | 1984 Dec. 8 | 3 | Ne | 800 | 30397 - 45035 | 0.029 | 11 | Solar Blind PMT |
| 4 | 1984 Dec. 9 | 1 | Ar | 400 | 30397 - 45035 | 0.029 | 8 | Solar Blind PMT |
| 5 | 1984 Dec. 9 | 2 | Ar | 200 | 30397 - 45035 | 0.029 | 8 | Solar Blind PMT |
| 6 | 1984 Dec. 9 | 8 | Ar | 725 | 30892 - 46247 | 0.029 | 30 | Solar Blind PMT |
| 7 | 1985 Aug. 30 | 1 | Ar | 600 | 31696 - 44799 | 0.025 | 4 | Solar Blind PMT |
| 8 | 1985 Aug. 30 | 2 | Ar | 920 | 31696 - 44799 | 0.025 | 8 | Solar Blind PMT |
| 9 | 1985 Aug. 30 | 3 | Ne | 1100 | 31637 - 44858 | 0.029 | 8 | Solar Blind PMT |
| 10 | 1988 Mar. 25 | 25 | Ar | 1500 | 30016 - 44848 | 0.029 | 7 | R166 PMT |
| spectra for study of the near-UV and blue multiplets of the $z^4D^o$ term[d] | | | | | | | | |
| 11 | 1989 Mar. 16 | 56 | Ne | 1300 | 7784 - 42809 | 0.050 | 2 | Super Blue Si PD |
| 12 | 1989 Mar. 16 | 57 | Ar | 844 | 7784 - 42809 | 0.050 | 6 | Super Blue Si PD |
| 13 | 1989 Feb. 26 | 2 | Ne | 1300 | 8291 - 43148 | 0.050 | 4 | Super Blue Si PD |
| 14 | 1989 Feb. 26 | 7 | Ne | 1300 | 8291 - 43317 | 0.050 | 4 | Super Blue Si PD |
| 15 | 1989 Feb. 26 | 14 | Ne | 1300 | 8291 - 43317 | 0.050 | 7 | Super Blue Si PD |
| 16 | 1989 Feb. 26 | 15 | Ar | 800 | 8291 - 43317 | 0.050 | 5 | Super Blue Si PD |
| 17 | 1989 Mar. 16 | 35 | Ne | 1300 | 7784 - 42809 | 0.050 | 6 | Super Blue Si PD |
| 18 | 1989 Mar. 16 | 42 | Ne | 1300 | 7784 - 42809 | 0.050 | 6 | Super Blue Si PD |
| 19 | 1989 Mar. 16 | 51 | Ne | 1300 | 7784 - 42809 | 0.050 | 6 | Super Blue Si PD |
| 20 | 1982 Jun.25 | 4 | Ar | 500 per anode | 7664 - 44591 | 0.057 | 9 | UV PD with CS9-54 filter |
| 21 | 1983 Feb. 11 | 2 | Ne/Ar | 790 | 7999 - 49942 | 0.057 | 8 | Midrange Si PD |
| 22 | 1983 Feb. 26 | 5 | Ar | 1000 | 7908 - 49942 | 0.064 | 8 | Large PD |
| 23 | 1983 Nov. 15 | 6 | Ne/Ar | 800 | 7958 - 49687 | 0.057 | 8 | Midrange Si PD |
| 24 | 1984 Dec. 10 | 20 | Ar | 750 | 5930 - 36879 | 0.041 | 2 | Super Blue Si PD with WG295 |
| 25 | 1984 Dec. 11 | 21 | Ne/Ar | 420 | 5930 - 36879 | 0.041 | 5 | Super Blue Si PD with WG295 |
| 26 | 1984 Dec. 11 | 36 | Ne/Ar | 1155 | 5927 - 36866 | 0.041 | 2 | Super Blue Si PD with WG295 |
| 27 | 1985 Jul. 30 | 5 | Ar | 810 | 7166 - 45149 | 0.064 | 7 | Midrange Si PD |

Notes:

[a] All spectra were recorded using the 1 m FTS on the McMath-Pierce Solar telescope at the National Solar Observatory, Kitt Peak, AZ. The FTS was operated with a UV beam splitter.

[b] The detectors for the first 10 spectra are all described as Solar Blind Photomultipliers (PMTs), but only for index number 10 is it specifically identified as R166.

[c] Spectra with index numbers 1-10 were used to determine the relative strength of lines in the deep UV and separately near UV multiplets from the $z^4P^o$ levels.

[d] Spectra with index numbers 11-27 were used to determine the relative strength of lines in the near UV multiplet to that of the lines in the optical blue multiplet from the $z^4D^o$ levels.

Table 3

Potential blends with Fe II lines of interest considered in this study

| Fe II upper level | | | blended | possible blending partner upper level | | | | |
|---|---|---|---|---|---|---|---|---|
| term | $E_k$ ($cm^{-1}$) | clean lines ( Å ) | line (Å) | species | term | $E_k$ ($cm^{-1}$) | clean lines ( Å ) | comment |
| $z^6F^o_{7/2}$ | 42237.06 | N/A | 2388.63 | Fe II | $3d^6(^3P_2)4d\ ^2P_{3/2}$ | 105317.40 | N/A | a |
| $z^6D^o_{5/2}$ | 38858.97 | N/A | 2617.62 | Fe II | $3d^6(^3H)4d\ ^4G_{9/2}$ | 103771.34 | N/A | a |
| $z^6P^o_{5/2}$ | 43238.61 | 2332.80, 2348.30 | 2359.11 | Fe II | $3d^6\ (^5D)4d\ e^6G_{7/2}$ | 84710.75 | $e^6G_{9/2}$: 2357.05<br>$e^6G_{9/2}$: 2363.86 | b |
| $z^6P^o_{3/2}$ | 43620.98 | 2327.40, 2344.28 | 2338.01 | Fe II | $3d^6(^3F_2)4d\ ^4F_{9/2}$ | 104916.55 | 231.461, 2333.91 | c |
| $z^4D^o_{5/2}$ | 44784.79 | 2714.41, 2768.93 | 2746.98 | Fe I | $3d^6(^3H)4s4p(^3P^o)\ z^5H^o_6$ | 43321.10 | $z^5H^o_5$: 2772.07<br>$z^5H^o_4$: 2797.78<br>$z^5H^o_3$: 2828.81 | a,d |
| $z^4D^o_{1/2}$ | 45206.47 | 2375.19, 2749.49 | 2736.97 | Fe I | $3d^7(^4P)4p\ y^5S^o_2$ | 44511.81 | 2717.79, 3707.92 | e |
| $z^4P^o_{5/2}$ | 47389.81 | N/A | 2947.66 | Fe II | $3d^6\ (^3H)4d\ ^2H_{11/2}$ | 106045.71 | N/A | a |
| $z^4P^o_{5/2}$ | 46967.48 | N/A | 3002.64 | Fe II | $3d^6\ (^3H)5s\ e^2G_{9/2}$ | 103608.92 | N/A | f |

NOTES:

[a]These lines are listed as blends in the NIST ASD. However we see no measurable branching ratio difference between spectra using neon and those using argon buffer gas. These lines are not blended in our spectra. N/A - not applicable

[b]This line is listed as a blend in the NIST ASD. This blend is real and present in our spectra. The upper level of the blending partner has no other clean transitions to compare to. Instead, ratios are taken between the blended line and clean transitions from the J=9/2 level in the same term. A least-squares optimization is performed to find the blending fraction in each spectrum as described in the text. The line of interest is found to be 63% and 61% of the feature in the 15 mA and 20 mA neon spectra, respectively. It is found to be 72% and 71% of the feature in the 15 mA and 20 mA argon spectra, respectively.

[c]This line is not listed as a blend in the NIST ASD but yields measurably different branching ratios in neon and argon echelle spectra. Our spectrum analysis software looks for all possible blend transitions that obey parity and ΔJ selection rules for Fe I, Fe II and the first and second spectra of the buffer gas. This line has potential blends with two transitions of Fe II, 104916.55 – 62158.13 $cm^{-1}$ as well as 113073.18 – 70314.62 $cm^{-1}$. The level at 104916.55 $cm^{-1}$ has other transitions that are observed only on the neon spectra and it is reasonable to assume that this is the blending partner. Since the upper level of the blending partner is only populated in the neon discharge, one could determine the branching fraction exclusively from the argon spectra. We have chosen to utilize all our spectra by performing a least squares optimization to determine the blending fraction in the neon spectra. The line of interest is found to be 86% and 85% in the 15 mA and 20 mA neon spectra, respectively.

[d]This line is listed as a blend in the NIST ASD. The upper level of the neutral blending partner, $z^5H^o_6$, has no other transitions to compare to. Bergeson et al. (1996) concluded that this transition might well be a significant fraction of the observed line strength, and chose not to report branching fractions for the $z^4D^o_{1/2}$ level, although they did report branching ratios for the remaining lines. We compared the potential blend to clean transitions from other levels in the $z^5H^o$ term using a least-squares optimization. We determine the ion fraction of this line in our spectra to be >98%.

[e]This line is listed as a blend in the NIST ASD. This is a real blend with the neutral line indicated, and is separated by least-squares optimization. The line of interest is found to be 79% and 81% of the feature in the 15 and 20 mA neon spectra, respectively. It is found to be 87% and 88% of the feature in the 15 and 20 mA argon spectra, respectively.

[f]This line is listed as a blend in the NIST ASD. This blend is real and was separated by the center-of-gravity technique on the FTS spectra as described in the text. N/A - not applicable

Table 4

Branching Fractions in Fe II

| Upper Level | | Lower Level | | $\lambda_{air}$ | Branching Fractions | | | | | |
|---|---|---|---|---|---|---|---|---|---|---|
| Term | $E_k$ (cm$^{-1}$) | Term | $E_i$ (cm$^{-1}$) | (Å) | This exp. | (±%) | Berg96[a] | (±%) | SSK04[b] | (±%) |
| z $^6D^o_{9/2}$ | 38458.99 | a $^6D_{9/2}$ | 0.00 | 2599.395 | 0.869 | (0.5) | 0.872 | (0.6) | 0.865 | (6) |
| | | a $^6D_{7/2}$ | 384.79 | 2625.667 | 0.129 | (3) | 0.126 | (4) | 0.134 | (6) |
| | | a $^4F_{9/2}$ | 1872.60 | 2732.448 | 0.000280 | (8) | . . . | . . . | 0.000362 | (8) |
| | | a $^4D_{7/2}$ | 7955.32 | 3277.349 | 0.00167 | (20) | . . . | . . . | 0.00122 | (8) |
| z $^6D^o_{7/2}$ | 38660.05 | a $^6D_{9/2}$ | 0.00 | 2585.876 | 0.320 | (1) | 0.317 | (3) | 0.340 | (6) |
| | | a $^6D_{7/2}$ | 384.79 | 2611.873 | 0.459 | (1) | 0.453 | (2) | 0.425 | (6) |
| | | a $^6D_{5/2}$ | 667.68 | 2631.323 | 0.220 | (2) | 0.229 | (3) | 0.234 | (6) |
| | | a $^4D_{7/2}$ | 7955.32 | 3255.888 | 0.00089 | (10) | . . . | . . . | 0.00102 | (8) |
| z $^6D^o_{5/2}$ | 38858.97 | a $^6D_{7/2}$ | 384.79 | 2598.369 | 0.524 | (1) | 0.516 | (2) | 0.528 | (6) |
| | | a $^6D_{5/2}$ | 667.68 | 2617.617 | 0.178 | (1) | 0.178 | (4) | 0.179 | (8) |
| | | a $^6D_{3/2}$ | 862.61 | 2631.047 | 0.299 | (1) | 0.305 | (4) | 0.292 | (6) |
| z $^6D^o_{3/2}$ | 39013.22 | a $^6D_{5/2}$ | 667.68 | 2607.087 | 0.657 | (1) | 0.665 | (2) | 0.642 | (6) |
| | | a $^6D_{3/2}$ | 862.61 | 2620.409 | 0.0130 | (3) | 0.0146 | (14) | 0.0160 | (8) |
| | | a $^6D_{1/2}$ | 977.05 | 2628.293 | 0.329 | (2) | 0.319 | (3) | 0.341 | (6) |
| z $^6D^o_{1/2}$ | 39109.32 | a $^6D_{3/2}$ | 862.61 | 2613.824 | 0.798 | (0.5) | 0.793 | (1) | 0.785 | (6) |
| | | a $^6D_{1/2}$ | 977.05 | 2621.669 | 0.199 | (1) | 0.205 | (4) | 0.212 | (6) |
| z $^6F^o_{11/2}$ | 41968.07 | a $^6D_{9/2}$ | 0.00 | 2382.037 | 1.000 | (0.1) | 1.00 | . . . | 1 | (2) |
| z $^6F^o_{9/2}$ | 42114.84 | a $^6D_{9/2}$ | 0.00 | 2373.735 | 0.120 | (2) | 0.120 | (4) | 0.158 | (6) |
| | | a $^6D_{7/2}$ | 384.79 | 2395.625 | 0.863 | (0.5) | 0.864 | (0.6) | 0.822 | (6) |
| | | a $^4D_{7/2}$ | 7955.32 | 2926.585 | 0.0134 | (8) | . . . | . . . | 0.0168 | (24) |
| z $^6F^o_{7/2}$ | 42237.06 | a $^6D_{7/2}$ | 384.79 | 2388.629 | 0.318 | (1) | 0.314 | (4) | 0.371 | (6) |
| | | a $^6D_{5/2}$ | 667.68 | 2404.885 | 0.665 | (0.5) | 0.668 | (2) | 0.612 | (6) |
| | | a $^4D_{5/2}$ | 8391.96 | 2953.774 | 0.0130 | (8) | . . . | . . . | 0.0169 | (24) |
| z $^6F^o_{5/2}$ | 42334.84 | a $^6D_{7/2}$ | 384.79 | 2383.06 | 0.0258 | (2) | . . . | . . . | . . . | . . . |
| | | a $^6D_{5/2}$ | 667.68 | 2399.241 | 0.453 | (1) | 0.458 | (2) | 0.462 | (6) |
| | | a $^6D_{3/2}$ | 862.61 | 2410.519 | 0.507 | (1) | 0.532 | (2) | 0.495 | (6) |
| | | a $^4D_{3/2}$ | 8680.47 | 2970.515 | 0.0093 | (8) | . . . | . . . | 0.00892 | (8) |
| z $^6F^o_{3/2}$ | 42401.32 | a $^6D_{5/2}$ | 667.68 | 2395.419 | 0.085 | (3) | 0.089 | (6) | 0.108 | (24) |
| | | a $^6D_{3/2}$ | 862.61 | 2406.661 | 0.568 | (0.5) | 0.568 | (2) | 0.523 | (6) |
| | | a $^6D_{1/2}$ | 977.05 | 2413.31 | 0.337 | (1) | 0.335 | (4) | 0.359 | (6) |
| | | a $^4F_{5/2}$ | 2837.98 | 2526.833 | 0.00231 | (9) | . . . | . . . | . . . | . . . |
| | | a $^4D_{5/2}$ | 8391.96 | 2939.507 | 0.00136 | (11) | . . . | . . . | . . . | . . . |
| | | a $^4D_{1/2}$ | 8846.78 | 2979.353 | 0.00615 | (9) | . . . | . . . | 0.00555 | (24) |
| z $^6F^o_{1/2}$ | 42439.85 | a $^6D_{3/2}$ | 862.61 | 2404.43 | 0.194 | (1) | 0.199 | (4) | 0.227 | (6) |
| | | a $^6D_{1/2}$ | 977.05 | 2411.067 | 0.797 | (0.5) | 0.794 | (1) | 0.768 | (6) |
| | | a $^4D_{3/2}$ | 8680.47 | 2961.275 | 0.00295 | (16) | . . . | . . . | . . . | . . . |
| | | a $^4D_{1/2}$ | 8846.78 | 2975.936 | 0.00323 | (13) | . . . | . . . | . . . | . . . |

| | | | | | | | |
|---|---|---|---|---|---|---|---|
| z $^{6}P^{o}_{7/2}$ | 42658.24 | a $^{6}D_{9/2}$ | 0.00 | 2343.495 | 0.644 (0.5) | 0.644 (1) | 0.63 (11) |
| | | a $^{6}D_{7/2}$ | 384.79 | 2364.828 | 0.216 (1) | 0.220 (4) | 0.224 (11) |
| | | a $^{6}D_{5/2}$ | 667.68 | 2380.761 | 0.113 (1) | 0.111 (6) | 0.119 (11) |
| | | a $^{4}F_{9/2}$ | 1872.60 | 2451.101 | 0.00120 (6) | . . . . . . | . . . . . . |
| | | a $^{4}D_{7/2}$ | 7955.32 | 2880.756 | 0.0084 (9) | . . . . . . | 0.00820 (23) |
| | | a $^{4}D_{5/2}$ | 8391.96 | 2917.466 | 0.00067 (13) | . . . . . . | . . . . . . |
| z $^{6}P^{o}_{5/2}$ | 43238.61 | a $^{6}D_{7/2}$ | 384.79 | 2332.799 | 0.438 (1) | . . . . . . | . . . . . . |
| | | a $^{6}D_{5/2}$ | 667.68 | 2348.302 | 0.384 (1) | . . . . . . | . . . . . . |
| | | a $^{6}D_{3/2}$ | 862.61 | 2359.105 | 0.164 (1) | . . . . . . | . . . . . . |
| | | a $^{4}D_{5/2}$ | 8391.96 | 2868.874 | 0.00318 (9) | . . . . . . | . . . . . . |
| z $^{6}P^{o}_{3/2}$ | 43620.98 | a $^{6}D_{5/2}$ | 667.68 | 2327.395 | 0.236 (1) | 0.236 (3) | 0.218 (11) |
| | | a $^{6}D_{3/2}$ | 862.61 | 2338.007 | 0.404 (1) | 0.406 (2) | 0.436 (11) |
| | | a $^{6}D_{1/2}$ | 977.05 | 2344.281 | 0.345 (1) | 0.344 (1) | 0.333 (11) |
| z $^{4}F^{o}_{9/2}$ | 44232.54 | a $^{6}D_{9/2}$ | 0.00 | 2260.079 | 0.0098 (9) | 0.0124 (6) | 0.0158 (24) |
| | | a $^{6}D_{7/2}$ | 384.79 | 2279.915 | 0.0163 (8) | 0.0174 (2) | . . . . . . |
| | | a $^{4}F_{9/2}$ | 1872.60 | 2359.999 | 0.113 (6) | 0.127 (6) | 0.145 (6) |
| | | a $^{4}F_{7/2}$ | 2430.14 | 2391.478 | 0.0113 (6) | 0.0142 (6) | 0.0145 (8) |
| | | a $^{4}D_{7/2}$ | 7955.32 | 2755.736 | 0.847 (1) | 0.824 (1) | 0.809 (6) |
| z $^{4}F^{o}_{7/2}$ | 44753.82 | a $^{6}D_{7/2}$ | 384.79 | 2253.126 | 0.0125 (9) | . . . . . . | 0.0158 (13) |
| | | a $^{6}D_{5/2}$ | 667.68 | 2267.586 | 0.0104 (9) | 0.0128 (9) | 0.0139 (13) |
| | | a $^{4}F_{9/2}$ | 1872.60 | 2331.308 | 0.099 (6) | 0.115 (5) | 0.124 (13) |
| | | a $^{4}F_{7/2}$ | 2430.14 | 2362.021 | 0.0449 (6) | 0.048 (8) | 0.0541 (13) |
| | | a $^{4}F_{5/2}$ | 2837.98 | 2385.006 | 0.0105 (6) | 0.0121 (10) | 0.0139 (13) |
| | | a $^{4}D_{5/2}$ | 8391.96 | 2749.321 | 0.813 (1) | 0.783 (1) | 0.772 (13) |
| | | a $^{4}P_{5/2}$ | 13474.45 | 3196.071 | 0.00532 (6) | . . . . . . | 0.00579 (24) |
| z $^{4}F^{o}_{5/2}$ | 45079.90 | a $^{6}D_{5/2}$ | 667.68 | 2250.935 | 0.0093 (8) | 0.0109 (10) | 0.0124 (13) |
| | | a $^{6}D_{3/2}$ | 862.61 | 2260.859 | 0.0065 (9) | 0.0074 (11) | 0.0084 (13) |
| | | a $^{4}F_{7/2}$ | 2430.14 | 2343.961 | 0.107 (6) | 0.117 (5) | 0.120 (13) |
| | | a $^{4}F_{5/2}$ | 2837.98 | 2366.594 | 0.0350 (6) | 0.0376 (6) | 0.0400 (13) |
| | | a $^{4}F_{3/2}$ | 3117.49 | 2382.358 | 0.0096 (7) | 0.0120 (10) | 0.0116 (24) |
| | | a $^{4}D_{7/2}$ | 7955.32 | 2692.834 | 0.00544 (3) | 0.0061 (13) | . . . . . . |
| | | a $^{4}D_{5/2}$ | 8391.96 | 2724.884 | 0.0352 (2) | 0.0361 (6) | 0.0376 (24) |
| | | a $^{4}D_{3/2}$ | 8680.47 | 2746.483 | 0.785 (1) | 0.767 (1) | 0.760 (13) |
| | | a $^{4}P_{3/2}$ | 13673.20 | 3183.114 | 0.00389 (8) | . . . . . . | 0.00348 (13) |
| z $^{4}F^{o}_{3/2}$ | 45289.82 | a $^{6}D_{3/2}$ | 862.61 | 2250.175 | 0.00515 (13) | . . . . . . | 0.00617 (11) |
| | | a $^{6}D_{1/2}$ | 977.05 | 2255.987 | 0.00255 (12) | . . . . . . | . . . . . . |
| | | a $^{4}F_{5/2}$ | 2837.98 | 2354.89 | 0.094 (6) | 0.099 (5) | 0.107 (11) |
| | | a $^{4}F_{3/2}$ | 3117.49 | 2370.499 | 0.0553 (6) | 0.058 (7) | 0.0700 (11) |
| | | a $^{4}D_{3/2}$ | 8680.47 | 2730.734 | 0.103 (2) | 0.103 (5) | 0.107 (11) |
| | | a $^{4}D_{1/2}$ | 8846.78 | 2743.197 | 0.733 (1) | 0.728 (1) | .700 (11) |
| z $^{4}D^{o}_{7/2}$ | 44446.91 | a $^{6}D_{9/2}$ | 0.00 | 2249.178 | 0.0077 (10) | 0.0091 (7) | 0.0158 (24) |
| | | a $^{6}D_{5/2}$ | 667.68 | 2283.484 | 0.00181 (11) | . . . . . . | . . . . . . |
| | | a $^{4}F_{9/2}$ | 1872.60 | 2348.116 | 0.186 (5) | 0.195 (5) | 0.211 (6) |
| | | a $^4F_{7/2}$ | 2430.14 | 2379.276 | 0.0646 (6) | 0.0733 (2) | 0.0902 (6) |
| | | a $^4F_{5/2}$ | 2837.98 | 2402.599 | 0.00634 (6) | 0.0065 (9) | 0.00940 (24) |
| | | a $^4D_{7/2}$ | 7955.32 | 2739.547 | 0.705 (1) | 0.682 (2) | 0.639 (6) |
| | | a $^4P_{5/2}$ | 13474.45 | 3227.743 | 0.0224 (6) | . . . . . . | 0.0263 (24) |
| | | b $^4P_{5/2}$ | 20830.55 | 4233.162 | 0.00134 (12) | . . . . . . | 0.00214 (14) |
| | | b $^4F_{9/2}$ | 22637.20 | 4583.829 | 0.00135 (14) | . . . . . . | 0.00214 (13) |
| z $^4D^o_{5/2}$ | 44784.79 | a $^6D_{7/2}$ | 384.79 | 2251.555 | 0.00213 (11) | . . . . . . | . . . . . . |
| | | a $^6D_{5/2}$ | 667.68 | 2265.994 | 0.00224 (11) | . . . . . . | . . . . . . |
| | | a $^4F_{7/2}$ | 2430.14 | 2360.294 | 0.156 (5) | . . . . . . | 0.181 (6) |
| | | a $^4F_{5/2}$ | 2837.98 | 2383.245 | 0.089 (5) | . . . . . . | 0.104 (6) |
| | | a $^4D_{7/2}$ | 7955.32 | 2714.413 | 0.166 (1) | . . . . . . | 0.165 (6) |
| | | a $^4D_{5/2}$ | 8391.96 | 2746.982 | 0.537 (1) | . . . . . . | 0.490 (6) |
| | | a $^4D_{3/2}$ | 8680.47 | 2768.934 | 0.0133 (3) | . . . . . . | 0.0138 (8) |
| | | a $^4P_{5/2}$ | 13474.45 | 3192.91 | 0.00415 (9) | . . . . . . | 0.00367 (24) |
| | | a $^4P_{3/2}$ | 13673.20 | 3213.309 | 0.0186 (6) | . . . . . . | 0.0178 (13) |
| | | b $^4P_{5/2}$ | 20830.55 | 4173.451 | 0.00080 (11) | . . . . . . | 0.00129 (13) |
| | | b $^4P_{3/2}$ | 21812.05 | 4351.762 | 0.00198 (17) | . . . . . . | 0.00141 (14) |
| | | b $^4F_{7/2}$ | 22810.35 | 4549.466 | 0.00193 (15) | . . . . . . | 0.00291 (13) |
| z $^4D^o_{3/2}$ | 45044.19 | a $^6D_{3/2}$ | 862.61 | 2262.687 | 0.0054 (10) | . . . . . . | 0.00576 (8) |
| | | a $^4F_{5/2}$ | 2837.98 | 2368.596 | 0.177 (5) | 0.181 (4) | 0.189 (6) |
| | | a $^4F_{3/2}$ | 3117.49 | 2384.388 | 0.095 (6) | 0.097 (5) | 0.112 (24) |
| | | a $^4D_{5/2}$ | 8391.96 | 2727.539 | 0.284 (1) | 0.284 (3) | 0.272 (6) |
| | | a $^4D_{3/2}$ | 8680.47 | 2749.181 | 0.373 (1) | 0.364 (2) | 0.352 (6) |
| | | a $^4D_{1/2}$ | 8846.78 | 2761.813 | 0.0397 (2) | 0.0412 (4) | 0.0448 (24) |
| | | a $^4P_{3/2}$ | 13673.20 | 3186.737 | 0.0110 (6) | . . . . . . | 0.0112 (8) |
| | | a $^4P_{1/2}$ | 13904.86 | 3210.445 | 0.0112 (7) | . . . . . . | 0.0106 (8) |
| | | b $^4P_{3/2}$ | 21812.05 | 4303.170 | 0.00079 (27) | . . . . . . | 0.000640 (26) |
| | | b $^4P_{1/2}$ | 22409.82 | 4416.819 | 0.00067 (22) | . . . . . . | . . . . . . |
| | | b $^4F_{5/2}$ | 22939.35 | 4522.628 | 0.00122 (13) | . . . . . . | 0.00243 (24) |
| z $^4D^o_{1/2}$ | 45206.47 | a $^4F_{3/2}$ | 3117.49 | 2375.194 | 0.275 (4) | 0.285 (4) | . . . . . . |
| | | a $^4D_{3/2}$ | 8680.47 | 2736.966 | 0.350 (2) | 0.353 (3) | . . . . . . |
| | | a $^4D_{1/2}$ | 8846.78 | 2749.486 | 0.340 (2) | 0.336 (2) | . . . . . . |
| | | a $^4P_{3/2}$ | 13673.20 | 3170.337 | 0.0040 (16) | . . . . . . | . . . . . . |
| | | a $^4P_{1/2}$ | 13904.86 | 3193.801 | 0.0211 (6) | . . . . . . | . . . . . . |
| | | b $^4P_{1/2}$ | 22409.82 | 4385.377 | 0.00116 (15) | . . . . . . | . . . . . . |
| | | b $^4F_{3/2}$ | 23031.28 | 4508.28 | 0.00183 (18) | . . . . . . | . . . . . . |
| z $^4P^o_{5/2}$ | 46967.48 | a $^4D_{7/2}$ | 7955.32 | 2562.535 | 0.593 (1) | . . . . . . | 0.585 (13) |
| | | a $^4D_{5/2}$ | 8391.96 | 2591.543 | 0.178 (2) | . . . . . . | 0.187 (13) |
| | | a $^4D_{3/2}$ | 8680.47 | 2611.073 | 0.0234 (3) | . . . . . . | 0.0238 (13) |
| | | a $^4P_{5/2}$ | 13474.45 | 2984.825 | 0.145 (5) | . . . . . . | 0.140 (24) |
| | | a $^4P_{3/2}$ | 13673.20 | 3002.644 | 0.055 (10) | . . . . . . | 0.0585 (13) |
| z $^4P^o_{3/2}$ | 47389.81 | a $^4D_{5/2}$ | 8391.96 | 2563.475 | 0.468 (1) | . . . . . . | 0.487 (6) |
| | | a $^4D_{3/2}$ | 8680.47 | 2582.583 | 0.274 (1) | . . . . . . | 0.284 (6) |
| | | a $^4D_{1/2}$ | 8846.78 | 2593.728 | 0.0483 (2) | . . . . . . | 0.0528 (8) |
| | | a $^4P_{5/2}$ | 13474.45 | 2947.655 | 0.097 (5) | . . . . . . | 0.0650 (8) |
| | | a $^4P_{3/2}$ | 13673.20 | 2965.032 | 0.0278 (6) | . . . . . . | 0.0304 (8) |
| | | a $^4P_{1/2}$ | 13904.86 | 2985.545 | 0.078 (6) | . . . . . . | 0.0771 (8) |
| z $^4P^o_{1/2}$ | 47626.11 | a $^4D_{3/2}$ | 8680.47 | 2566.912 | 0.379 (1) | . . . . . . | . . . . . . |
| | | a $^4D_{1/2}$ | 8846.78 | 2577.922 | 0.407 (1) | . . . . . . | . . . . . . |
| | | a $^4P_{3/2}$ | 13673.20 | 2944.395 | 0.178 (4) | . . . . . . | . . . . . . |
| | | a $^4P_{1/2}$ | 13904.86 | 2964.623 | 0.0307 (8) | . . . . . . | . . . . . . |

NOTES: [a] Bergeson et al. (1996), [b] Schnabel et al. (2004).

Table 5
Radiative lifetimes used in this study

| level | lifetime (ns) | | | ref(s). |
|---|---|---|---|---|
| $z^6D^o_{9/2}$ | 3.69 | ± | 0.07 | a,b |
| $z^6D^o_{7/2}$ | 3.68 | ± | 0.08 | a,b |
| $z^6D^o_{5/2}$ | 3.66 | ± | 0.07 | a,b |
| $z^6D^o_{3/2}$ | 3.78 | ± | 0.09 | a,b |
| $z^6D^o_{1/2}$ | 3.72 | ± | 0.11 | a,b |
| $z^6F^o_{11/2}$ | 3.20 | ± | 0.05 | a,b |
| $z^6F^o_{9/2}$ | 3.26 | ± | 0.05 | a,b |
| $z^6F^o_{7/2}$ | 3.26 | ± | 0.08 | a,b |
| $z^6F^o_{5/2}$ | 3.32 | ± | 0.07 | a,b |
| $z^6F^o_{3/2}$ | 3.40 | ± | 0.11 | a,b |
| $z^6F^o_{1/2}$ | 3.30 | ± | 0.30 | c |
| $z^6P^o_{7/2}$ | 3.72 | ± | 0.05 | a,b,d |
| $z^6P^o_{5/2}$ | 3.79 | ± | 0.10 | a,b,d |
| $z^6P^o_{3/2}$ | 3.71 | ± | 0.12 | b,d |
| $z^4F^o_{9/2}$ | 3.80 | ± | 0.10 | b,d |
| $z^4F^o_{7/2}$ | 3.61 | ± | 0.11 | b,d |
| $z^4F^o_{5/2}$ | 3.65 | ± | 0.11 | b,d |
| $z^4F^o_{3/2}$ | 3.70 | ± | 0.20 | c |
| $z^4D^o_{7/2}$ | 3.00 | ± | 0.06 | b,d |
| $z^4D^o_{5/2}$ | 3.00 | ± | 0.07 | b,d |
| $z^4D^o_{3/2}$ | 2.91 | ± | 0.09 | b |
| $z^4D^o_{1/2}$ | 2.90 | ± | 0.20 | c |
| $z^4P^o_{5/2}$ | 3.25 | ± | 0.10 | b |
| $z^4P^o_{3/2}$ | 3.25 | ± | 0.10 | b |
| $z^4P^o_{1/2}$ | 3.25 | ± | 0.10 | b |

NOTES:
[a]Biémont et al. (1991)
[b]Schnabel et al. (2004)
[c]Hannaford et al. (1992)
[d]Guo et al. (1992)

Table 6
Experimental Atomic Transition Probabilities for 131 lines of Fe II

| $\lambda_{air}$[a] (Å) | Upper Level[a] $E_U$ ($cm^{-1}$) | $J_U$ | Lower Level[a] $E_L$ ($cm^{-1}$) | $J_L$ | A-value ($10^6$ $s^{-1}$) | | | log(*gf*) | | |
|---|---|---|---|---|---|---|---|---|---|---|
| 2249.178 | 44446.9051 | 3.5 | 0.0000 | 4.5 | 2.58 | ± | 0.26 | -1.81 | ± | 0.04 |
| 2250.175 | 45289.8248 | 1.5 | 862.6118 | 1.5 | 1.40 | ± | 0.20 | -2.37 | ± | 0.06 |
| 2250.935 | 45079.9019 | 2.5 | 667.6829 | 2.5 | 2.57 | ± | 0.22 | -1.93 | ± | 0.04 |
| 2251.555 | 44784.7859 | 2.5 | 384.7872 | 3.5 | 0.70 | ± | 0.08 | -2.49 | ± | 0.05 |
| 2253.126 | 44753.8179 | 3.5 | 384.7872 | 3.5 | 3.48 | ± | 0.33 | -1.68 | ± | 0.04 |
| 2255.987 | 45289.8248 | 1.5 | 977.0498 | 0.5 | 0.69 | ± | 0.09 | -2.68 | ± | 0.05 |
| 2260.079 | 44232.5398 | 4.5 | 0.0000 | 4.5 | 2.58 | ± | 0.24 | -1.70 | ± | 0.04 |
| 2260.859 | 45079.9019 | 2.5 | 862.6118 | 1.5 | 1.79 | ± | 0.17 | -2.09 | ± | 0.04 |
| 2262.687 | 45044.1916 | 1.5 | 862.6118 | 1.5 | 1.85 | ± | 0.19 | -2.25 | ± | 0.04 |
| 2265.994 | 44784.7859 | 2.5 | 667.6829 | 2.5 | 0.74 | ± | 0.08 | -2.46 | ± | 0.04 |
| 2267.586 | 44753.8179 | 3.5 | 667.6829 | 2.5 | 2.88 | ± | 0.27 | -1.75 | ± | 0.04 |
| 2279.915 | 44232.5398 | 4.5 | 384.7872 | 3.5 | 4.31 | ± | 0.36 | -1.48 | ± | 0.03 |
| 2283.484 | 44446.9051 | 3.5 | 667.6829 | 2.5 | 0.61 | ± | 0.07 | -2.42 | ± | 0.05 |
| 2327.395 | 43620.9842 | 1.5 | 667.6829 | 2.5 | 63.5 | ± | 2.2 | -0.68 | ± | 0.01 |
| 2331.308 | 44753.8179 | 3.5 | 1872.5998 | 4.5 | 27.7 | ± | 1.8 | -0.75 | ± | 0.03 |
| 2332.799 | 43238.6066 | 2.5 | 384.7872 | 3.5 | 116.7 | ± | 3.3 | -0.25 | ± | 0.01 |
| 2338.007 | 43620.9842 | 1.5 | 862.6118 | 1.5 | 108.6 | ± | 3.7 | -0.45 | ± | 0.01 |
| 2343.495 | 42658.2444 | 3.5 | 0.0000 | 4.5 | 172.8 | ± | 2.4 | 0.06 | ± | 0.01 |
| 2343.961 | 45079.9019 | 2.5 | 2430.1369 | 3.5 | 29.2 | ± | 2.0 | -0.84 | ± | 0.03 |
| 2344.281 | 43620.9842 | 1.5 | 977.0498 | 0.5 | 92.8 | ± | 3.1 | -0.51 | ± | 0.01 |
| 2348.116 | 44446.9051 | 3.5 | 1872.5998 | 4.5 | 62.4 | ± | 3.3 | -0.39 | ± | 0.02 |
| 2348.302 | 43238.6066 | 2.5 | 667.6829 | 2.5 | 102.3 | ± | 2.9 | -0.30 | ± | 0.01 |
| 2354.889 | 45289.8248 | 1.5 | 2837.9807 | 2.5 | 25.5 | ± | 2.1 | -1.08 | ± | 0.03 |
| 2359.105 | 43238.6066 | 2.5 | 862.6118 | 1.5 | 43.8 | ± | 1.2 | -0.66 | ± | 0.01 |
| 2359.999 | 44232.5398 | 4.5 | 1872.5998 | 4.5 | 29.9 | ± | 1.9 | -0.60 | ± | 0.03 |
| 2360.294 | 44784.7859 | 2.5 | 2430.1369 | 3.5 | 51.7 | ± | 2.9 | -0.58 | ± | 0.02 |
| 2362.021 | 44753.8179 | 3.5 | 2430.1369 | 3.5 | 12.5 | ± | 0.8 | -1.08 | ± | 0.03 |
| 2364.828 | 42658.2444 | 3.5 | 384.7872 | 3.5 | 58.0 | ± | 1.0 | -0.41 | ± | 0.01 |
| 2366.594 | 45079.9019 | 2.5 | 2837.9807 | 2.5 | 9.6 | ± | 0.6 | -1.32 | ± | 0.03 |
| 2368.596 | 45044.1916 | 1.5 | 2837.9807 | 2.5 | 60.9 | ± | 3.6 | -0.69 | ± | 0.02 |
| 2370.499 | 45289.8248 | 1.5 | 3117.4877 | 1.5 | 15.1 | ± | 1.2 | -1.30 | ± | 0.03 |
| 2373.735 | 42114.8380 | 4.5 | 0.0000 | 4.5 | 36.9 | ± | 0.9 | -0.51 | ± | 0.01 |
| 2375.194 | 45206.4704 | 0.5 | 3117.4877 | 1.5 | 95 | ± | 8 | -0.79 | ± | 0.04 |
| 2379.276 | 44446.9051 | 3.5 | 2430.1369 | 3.5 | 21.7 | ± | 1.4 | -0.83 | ± | 0.03 |
| 2380.761 | 42658.2444 | 3.5 | 667.6829 | 2.5 | 30.4 | ± | 0.5 | -0.68 | ± | 0.01 |
| 2382.037 | 41968.0698 | 5.5 | 0.0000 | 4.5 | 313.0 | ± | 4.4 | 0.50 | ± | 0.01 |
| 2382.358 | 45079.9019 | 2.5 | 3117.4877 | 1.5 | 2.63 | ± | 0.20 | -1.87 | ± | 0.03 |
| 2383.060 | 42334.8444 | 2.5 | 384.7872 | 3.5 | 7.82 | ± | 0.23 | -1.40 | ± | 0.01 |
| 2383.245 | 44784.7859 | 2.5 | 2837.9807 | 2.5 | 29.5 | ± | 1.6 | -0.82 | ± | 0.02 |
| 2384.388 | 45044.1916 | 1.5 | 3117.4877 | 1.5 | 32.6 | ± | 2.2 | -0.96 | ± | 0.03 |
| 2385.006 | 44753.8179 | 3.5 | 2837.9807 | 2.5 | 2.92 | ± | 0.19 | -1.70 | ± | 0.03 |
| 2388.629 | 42237.0575 | 3.5 | 384.7872 | 3.5 | 98.0 | ± | 2.6 | -0.18 | ± | 0.01 |
| 2391.478 | 44232.5398 | 4.5 | 2430.1369 | 3.5 | 2.97 | ± | 0.19 | -1.59 | ± | 0.03 |
| 2395.419 | 42401.3198 | 1.5 | 667.6829 | 2.5 | 25.1 | ± | 1.1 | -1.06 | ± | 0.02 |
| 2395.625 | 42114.8380 | 4.5 | 384.7872 | 3.5 | 266 | ± | 4 | 0.36 | ± | 0.01 |
| 2399.241 | 42334.8444 | 2.5 | 667.6829 | 2.5 | 137.3 | ± | 3.3 | -0.15 | ± | 0.01 |
| 2402.599 | 44446.9051 | 3.5 | 2837.9807 | 2.5 | 2.13 | ± | 0.13 | -1.83 | ± | 0.03 |
| 2404.431 | 42439.8511 | 0.5 | 862.6118 | 1.5 | 59 | ± | 5 | -0.99 | ± | 0.04 |
| 2404.885 | 42237.0575 | 3.5 | 667.6829 | 2.5 | 205 | ± | 5 | 0.15 | ± | 0.01 |
| 2406.661 | 42401.3198 | 1.5 | 862.6118 | 1.5 | 167 | ± | 6 | -0.24 | ± | 0.02 |
| 2410.519 | 42334.8444 | 2.5 | 862.6118 | 1.5 | 153.7 | ± | 3.7 | -0.10 | ± | 0.01 |
| 2411.067 | 42439.8511 | 0.5 | 977.0498 | 0.5 | 242 | ± | 22 | -0.38 | ± | 0.04 |
| 2413.310 | 42401.3198 | 1.5 | 977.0498 | 0.5 | 99.2 | ± | 3.4 | -0.46 | ± | 0.01 |
| 2451.101 | 42658.2444 | 3.5 | 1872.5998 | 4.5 | 0.324 | ± | 0.020 | -2.63 | ± | 0.03 |
| 2526.833 | 42401.3198 | 1.5 | 2837.9807 | 2.5 | 0.68 | ± | 0.07 | -2.58 | ± | 0.04 |
| 2562.535 | 46967.4751 | 2.5 | 7955.3186 | 3.5 | 184 | ± | 6 | 0.03 | ± | 0.01 |
| 2563.475 | 47389.809 | 1.5 | 8391.9554 | 2.5 | 145 | ± | 5 | -0.25 | ± | 0.01 |
| 2566.912 | 47626.110 | 0.5 | 8680.4706 | 1.5 | 117.2 | ± | 3.8 | -0.64 | ± | 0.01 |
| 2577.922 | 47626.110 | 0.5 | 8846.7837 | 0.5 | 126 | ± | 4 | -0.60 | ± | 0.01 |
| 2582.583 | 47389.809 | 1.5 | 8680.4706 | 1.5 | 84.9 | ± | 2.7 | -0.47 | ± | 0.01 |
| 2585.876 | 38660.0537 | 3.5 | 0.0000 | 4.5 | 87.2 | ± | 2.1 | -0.16 | ± | 0.01 |
| 2591.543 | 46967.4751 | 2.5 | 8391.9554 | 2.5 | 55.1 | ± | 2.0 | -0.48 | ± | 0.02 |
| 2593.728 | 47389.809 | 1.5 | 8846.7837 | 0.5 | 14.9 | ± | 0.5 | -1.22 | ± | 0.01 |
| 2598.369 | 38858.9696 | 2.5 | 384.7872 | 3.5 | 143.1 | ± | 3.0 | -0.06 | ± | 0.01 |
| 2599.395 | 38458.9934 | 4.5 | 0.0000 | 4.5 | 236 | ± | 4 | 0.38 | ± | 0.01 |
| 2607.087 | 39013.2160 | 1.5 | 667.6829 | 2.5 | 174 | ± | 4 | -0.15 | ± | 0.01 |
| 2611.073 | 46967.4751 | 2.5 | 8680.4706 | 1.5 | 7.23 | ± | 0.31 | -1.36 | ± | 0.02 |
| 2611.873 | 38660.0537 | 3.5 | 384.7872 | 3.5 | 124.9 | ± | 3.0 | 0.01 | ± | 0.01 |
| 2613.824 | 39109.3161 | 0.5 | 862.6118 | 1.5 | 215 | ± | 6 | -0.36 | ± | 0.01 |
| 2617.617 | 38858.9696 | 2.5 | 667.6829 | 2.5 | 48.5 | ± | 1.0 | -0.52 | ± | 0.01 |
| 2620.409 | 39013.2160 | 1.5 | 862.6118 | 1.5 | 3.44 | ± | 0.13 | -1.85 | ± | 0.02 |
| 2621.669 | 39109.3161 | 0.5 | 977.0498 | 0.5 | 53.6 | ± | 1.6 | -0.96 | ± | 0.01 |
| 2625.667 | 38458.9934 | 4.5 | 384.7872 | 3.5 | 34.9 | ± | 1.2 | -0.44 | ± | 0.01 |
| 2628.293 | 39013.2160 | 1.5 | 977.0498 | 0.5 | 87.0 | ± | 2.6 | -0.44 | ± | 0.01 |
| 2631.047 | 38858.9696 | 2.5 | 862.6118 | 1.5 | 81.6 | ± | 1.7 | -0.29 | ± | 0.01 |
| 2631.323 | 38660.0537 | 3.5 | 667.6829 | 2.5 | 59.8 | ± | 1.8 | -0.30 | ± | 0.01 |
| 2692.834 | 45079.9019 | 2.5 | 7955.3186 | 3.5 | 1.50 | ± | 0.07 | -2.01 | ± | 0.02 |
| 2714.413 | 44784.7859 | 2.5 | 7955.3186 | 3.5 | 54.8 | ± | 1.4 | -0.44 | ± | 0.01 |
| 2724.884 | 45079.9019 | 2.5 | 8391.9554 | 2.5 | 9.69 | ± | 0.36 | -1.19 | ± | 0.02 |
| 2727.539 | 45044.1916 | 1.5 | 8391.9554 | 2.5 | 98.0 | ± | 3.2 | -0.36 | ± | 0.01 |
| 2730.734 | 45289.8248 | 1.5 | 8680.4706 | 1.5 | 28.0 | ± | 1.6 | -0.91 | ± | 0.02 |
| 2732.448 | 38458.9934 | 4.5 | 1872.5998 | 4.5 | 0.076 | ± | 0.006 | -3.07 | ± | 0.03 |
| 2736.966 | 45206.4704 | 0.5 | 8680.4706 | 1.5 | 121 | ± | 9 | -0.57 | ± | 0.03 |
| 2739.547 | 44446.9051 | 3.5 | 7955.3186 | 3.5 | 237 | ± | 5 | 0.33 | ± | 0.01 |
| 2743.197 | 45289.8248 | 1.5 | 8846.7837 | 0.5 | 200 | ± | 11 | -0.05 | ± | 0.02 |
| 2746.483 | 45079.9019 | 2.5 | 8680.4706 | 1.5 | 216 | ± | 7 | 0.16 | ± | 0.01 |
| 2746.982 | 44784.7859 | 2.5 | 8391.9554 | 2.5 | 178 | ± | 5 | 0.08 | ± | 0.01 |
| 2749.181 | 45044.1916 | 1.5 | 8680.4706 | 1.5 | 129 | ± | 4 | -0.24 | ± | 0.01 |
| 2749.321 | 44753.8179 | 3.5 | 8391.9554 | 2.5 | 226 | ± | 7 | 0.31 | ± | 0.01 |
| 2749.486 | 45206.4704 | 0.5 | 8846.7837 | 0.5 | 117 | ± | 8 | -0.58 | ± | 0.03 |
| 2755.736 | 44232.5398 | 4.5 | 7955.3186 | 3.5 | 224 | ± | 6 | 0.41 | ± | 0.01 |
| 2761.813 | 45044.1916 | 1.5 | 8846.7837 | 0.5 | 13.7 | ± | 0.5 | -1.20 | ± | 0.02 |
| 2768.934 | 44784.7859 | 2.5 | 8680.4706 | 1.5 | 4.42 | ± | 0.17 | -1.51 | ± | 0.02 |
| 2868.874 | 43238.6066 | 2.5 | 8391.9554 | 2.5 | 0.85 | ± | 0.08 | -2.21 | ± | 0.04 |
| 2880.756 | 42658.2444 | 3.5 | 7955.3186 | 3.5 | 2.26 | ± | 0.21 | -1.65 | ± | 0.04 |
| 2917.466 | 42658.2444 | 3.5 | 8391.9554 | 2.5 | 0.179 | ± | 0.023 | -2.74 | ± | 0.05 |
| 2926.585 | 42114.8380 | 4.5 | 7955.3186 | 3.5 | 4.13 | ± | 0.34 | -1.28 | ± | 0.03 |
| 2939.507 | 42401.3198 | 1.5 | 8391.9554 | 2.5 | 0.40 | ± | 0.05 | -2.68 | ± | 0.05 |
| 2944.395 | 47626.110 | 0.5 | 13673.2045 | 1.5 | 55.1 | ± | 2.8 | -0.85 | ± | 0.02 |
| 2947.655 | 47389.809 | 1.5 | 13474.4474 | 2.5 | 30.2 | ± | 1.8 | -0.81 | ± | 0.03 |
| 2953.774 | 42237.0575 | 3.5 | 8391.9554 | 2.5 | 4.00 | ± | 0.33 | -1.38 | ± | 0.03 |
| 2961.275 | 42439.8511 | 0.5 | 8680.4706 | 1.5 | 0.90 | ± | 0.17 | -2.63 | ± | 0.08 |
| 2964.623 | 47626.110 | 0.5 | 13904.8604 | 0.5 | 9.5 | ± | 0.8 | -1.60 | ± | 0.04 |
| 2965.032 | 47389.809 | 1.5 | 13673.2045 | 1.5 | 8.6 | ± | 0.6 | -1.35 | ± | 0.03 |
| 2970.515 | 42334.8444 | 2.5 | 8680.4706 | 1.5 | 2.81 | ± | 0.23 | -1.65 | ± | 0.03 |
| 2975.936 | 42439.8511 | 0.5 | 8846.7837 | 0.5 | 0.98 | ± | 0.16 | -2.59 | ± | 0.07 |
| 2979.353 | 42401.3198 | 1.5 | 8846.7837 | 0.5 | 1.81 | ± | 0.17 | -2.02 | ± | 0.04 |
| 2984.825 | 46967.4751 | 2.5 | 13474.4474 | 2.5 | 44.9 | ± | 2.6 | -0.45 | ± | 0.02 |
| 2985.545 | 47389.809 | 1.5 | 13904.8604 | 0.5 | 24.0 | ± | 1.6 | -0.89 | ± | 0.03 |
| 3002.644 | 46967.4751 | 2.5 | 13673.2045 | 1.5 | 16.9 | ± | 1.8 | -0.87 | ± | 0.04 |
| 3170.337 | 45206.4704 | 0.5 | 13673.2045 | 1.5 | 1.30 | ± | 0.23 | -2.38 | ± | 0.07 |
| 3183.114 | 45079.9019 | 2.5 | 13673.2045 | 1.5 | 1.07 | ± | 0.09 | -2.01 | ± | 0.04 |
| 3186.737 | 45044.1916 | 1.5 | 13673.2045 | 1.5 | 3.84 | ± | 0.26 | -1.64 | ± | 0.03 |
| 3192.910 | 44784.7859 | 2.5 | 13474.4474 | 2.5 | 1.36 | ± | 0.13 | -1.90 | ± | 0.04 |
| 3193.801 | 45206.4704 | 0.5 | 13904.8604 | 0.5 | 7.4 | ± | 0.7 | -1.65 | ± | 0.04 |
| 3196.071 | 44753.8179 | 3.5 | 13474.4474 | 2.5 | 1.48 | ± | 0.10 | -1.74 | ± | 0.03 |
| 3210.445 | 45044.1916 | 1.5 | 13904.8604 | 0.5 | 3.80 | ± | 0.29 | -1.62 | ± | 0.03 |
| 3213.309 | 44784.7859 | 2.5 | 13673.2045 | 1.5 | 6.2 | ± | 0.4 | -1.24 | ± | 0.03 |
| 3227.743 | 44446.9051 | 3.5 | 13474.4474 | 2.5 | 7.5 | ± | 0.5 | -1.03 | ± | 0.03 |
| 3255.888 | 38660.0537 | 3.5 | 7955.3186 | 3.5 | 0.242 | ± | 0.025 | -2.51 | ± | 0.04 |
| 3277.349 | 38458.9934 | 4.5 | 7955.3186 | 3.5 | 0.45 | ± | 0.09 | -2.14 | ± | 0.08 |
| 4173.451 | 44784.7859 | 2.5 | 20830.5534 | 2.5 | 0.268 | ± | 0.030 | -2.38 | ± | 0.05 |
| 4233.162 | 44446.9051 | 3.5 | 20830.5534 | 2.5 | 0.45 | ± | 0.05 | -2.02 | ± | 0.05 |
| 4303.170 | 45044.1916 | 1.5 | 21812.0454 | 1.5 | 0.27 | ± | 0.07 | -2.52 | ± | 0.10 |
| 4351.762 | 44784.7859 | 2.5 | 21812.0454 | 1.5 | 0.66 | ± | 0.11 | -1.95 | ± | 0.07 |
| 4385.377 | 45206.4704 | 0.5 | 22409.8178 | 0.5 | 0.40 | ± | 0.07 | -2.64 | ± | 0.07 |
| 4416.819 | 45044.1916 | 1.5 | 22409.8178 | 0.5 | 0.23 | ± | 0.05 | -2.57 | ± | 0.09 |
| 4508.281 | 45206.4704 | 0.5 | 23031.2829 | 1.5 | 0.63 | ± | 0.10 | -2.42 | ± | 0.06 |
| 4522.628 | 45044.1916 | 1.5 | 22939.3512 | 2.5 | 0.42 | ± | 0.06 | -2.29 | ± | 0.06 |
| 4549.466 | 44784.7859 | 2.5 | 22810.3459 | 3.5 | 0.64 | ± | 0.10 | -1.92 | ± | 0.06 |
| 4583.829 | 44446.9051 | 3.5 | 22637.1950 | 4.5 | 0.45 | ± | 0.06 | -1.94 | ± | 0.05 |

NOTES: [a] wavelengths and energy levels from Nave & Johansson (2013)

Title: ATOMIC TRANSITION PROBABILITIES FOR UV AND BLUE LINES OF Fe II AND ABUNDANCE DETERMINATIONS IN THE PHOTOSPHERES OF THE SUN AND METAL-POOR STAR HD 84937
Authors: Den Hartog, E.A., Lawler J.E., Sneden C., Cowan J.J., Brukhovesky, A.
Table: Experimental Atomic Transition Probabilities for 131 lines of Fe II
===================================================================================
Byte-by-byte Description of file: apjsxxxxt6_mrt.txt
-----------------------------------------------------------------------------------

| Bytes | Format | Units | Label | Explanations |
|---|---|---|---|---|
| 1- 8 | F8.3 | A | WaveAir | Wavelength in air; Angstroms (1) |
| 10- 19 | F10.4 | cm-1 | UpLev | Upper level energy (1) |
| 21- 23 | F3.1 | --- | UpJ | Upper level J (1) |
| 25- 34 | F10.4 | cm-1 | LowLev | Lower level energy (1) |
| 36- 38 | F3.1 | --- | LowJ | Lower level J (1) |
| 40- 46 | F7.3 | 10+6/s | TranP | Transition probability |
| 48- 53 | F6.3 | 10+6/s | e_TranP | Total uncertainty in TranP |
| 55- 59 | F5.2 | --- | log(gf) | Log of degeneracy times oscillator strength |
| 61- 64 | F4.2 | --- | e_log(gf) | uncertainty in log(gf) |

-----------------------------------------------------------------------------------
Note (1): wavelengths and energy levels from Nave & Johansson (2013)
-----------------------------------------------------------------------------------

| | | | | | | | | |
|---|---|---|---|---|---|---|---|---|
| 2249.178 | 44446.9051 | 3.5 | 0.0000 | 4.5 | 2.58 | 0.26 | -1.81 | 0.04 |
| 2250.175 | 45289.8248 | 1.5 | 862.6118 | 1.5 | 1.40 | 0.20 | -2.37 | 0.06 |
| 2250.935 | 45079.9019 | 2.5 | 667.6829 | 2.5 | 2.57 | 0.22 | -1.93 | 0.04 |
| 2251.555 | 44784.7859 | 2.5 | 384.7872 | 3.5 | 0.70 | 0.08 | -2.49 | 0.05 |
| 2253.126 | 44753.8179 | 3.5 | 384.7872 | 3.5 | 3.48 | 0.33 | -1.68 | 0.04 |
| 2255.987 | 45289.8248 | 1.5 | 977.0498 | 0.5 | 0.69 | 0.09 | -2.68 | 0.05 |
| 2260.079 | 44232.5398 | 4.5 | 0.0000 | 4.5 | 2.58 | 0.24 | -1.70 | 0.04 |
| 2260.859 | 45079.9019 | 2.5 | 862.6118 | 1.5 | 1.79 | 0.17 | -2.09 | 0.04 |
| 2262.687 | 45044.1916 | 1.5 | 862.6118 | 1.5 | 1.85 | 0.19 | -2.25 | 0.04 |
| 2265.994 | 44784.7859 | 2.5 | 667.6829 | 2.5 | 0.74 | 0.08 | -2.46 | 0.04 |
| 2267.586 | 44753.8179 | 3.5 | 667.6829 | 2.5 | 2.88 | 0.27 | -1.75 | 0.04 |
| 2279.915 | 44232.5398 | 4.5 | 384.7872 | 3.5 | 4.31 | 0.36 | -1.48 | 0.03 |
| 2283.484 | 44446.9051 | 3.5 | 667.6829 | 2.5 | 0.61 | 0.07 | -2.42 | 0.05 |
| 2327.395 | 43620.9842 | 1.5 | 667.6829 | 2.5 | 63.5 | 2.2 | -0.68 | 0.01 |
| 2331.308 | 44753.8179 | 3.5 | 1872.5998 | 4.5 | 27.7 | 1.8 | -0.75 | 0.03 |
| 2332.799 | 43238.6066 | 2.5 | 384.7872 | 3.5 | 116.7 | 3.3 | -0.25 | 0.01 |
| 2338.007 | 43620.9842 | 1.5 | 862.6118 | 1.5 | 108.6 | 3.7 | -0.45 | 0.01 |
| 2343.495 | 42658.2444 | 3.5 | 0.0000 | 4.5 | 172.8 | 2.4 | 0.06 | 0.01 |
| 2343.961 | 45079.9019 | 2.5 | 2430.1369 | 3.5 | 29.2 | 2.0 | -0.84 | 0.03 |
| 2344.281 | 43620.9842 | 1.5 | 977.0498 | 0.5 | 92.8 | 3.1 | -0.51 | 0.01 |
| 2348.116 | 44446.9051 | 3.5 | 1872.5998 | 4.5 | 62.4 | 3.3 | -0.39 | 0.02 |
| 2348.302 | 43238.6066 | 2.5 | 667.6829 | 2.5 | 102.3 | 2.9 | -0.30 | 0.01 |
| 2354.889 | 45289.8248 | 1.5 | 2837.9807 | 2.5 | 25.5 | 2.1 | -1.08 | 0.03 |
| 2359.105 | 43238.6066 | 2.5 | 862.6118 | 1.5 | 43.8 | 1.2 | -0.66 | 0.01 |
| 2359.999 | 44232.5398 | 4.5 | 1872.5998 | 4.5 | 29.9 | 1.9 | -0.60 | 0.03 |
| 2360.294 | 44784.7859 | 2.5 | 2430.1369 | 3.5 | 51.7 | 2.9 | -0.58 | 0.02 |
| 2362.021 | 44753.8179 | 3.5 | 2430.1369 | 3.5 | 12.5 | 0.8 | -1.08 | 0.03 |
| 2364.828 | 42658.2444 | 3.5 | 384.7872 | 3.5 | 58.0 | 1.0 | -0.41 | 0.01 |
| 2366.594 | 45079.9019 | 2.5 | 2837.9807 | 2.5 | 9.6 | 0.6 | -1.32 | 0.03 |
| 2368.596 | 45044.1916 | 1.5 | 2837.9807 | 2.5 | 60.9 | 3.6 | -0.69 | 0.02 |
| 2370.499 | 45289.8248 | 1.5 | 3117.4877 | 1.5 | 15.1 | 1.2 | -1.30 | 0.03 |
| 2373.735 | 42114.8380 | 4.5 | 0.0000 | 4.5 | 36.9 | 0.9 | -0.51 | 0.01 |
| 2375.194 | 45206.4704 | 0.5 | 3117.4877 | 1.5 | 95. | 8. | -0.79 | 0.04 |
| 2379.276 | 44446.9051 | 3.5 | 2430.1369 | 3.5 | 21.7 | 1.4 | -0.83 | 0.03 |
| 2380.761 | 42658.2444 | 3.5 | 667.6829 | 2.5 | 30.4 | 0.5 | -0.68 | 0.01 |
| 2382.037 | 41968.0698 | 5.5 | 0.0000 | 4.5 | 313.0 | 4.4 | 0.50 | 0.01 |
| 2382.358 | 45079.9019 | 2.5 | 3117.4877 | 1.5 | 2.63 | 0.20 | -1.87 | 0.03 |
| 2383.060 | 42334.8444 | 2.5 | 384.7872 | 3.5 | 7.82 | 0.23 | -1.40 | 0.01 |
| 2383.245 | 44784.7859 | 2.5 | 2837.9807 | 2.5 | 29.5 | 1.6 | -0.82 | 0.02 |
| 2384.388 | 45044.1916 | 1.5 | 3117.4877 | 1.5 | 32.6 | 2.2 | -0.96 | 0.03 |
| 2385.006 | 44753.8179 | 3.5 | 2837.9807 | 2.5 | 2.92 | 0.19 | -1.70 | 0.03 |
| 2388.629 | 42237.0575 | 3.5 | 384.7872 | 3.5 | 98.0 | 2.6 | -0.18 | 0.01 |
| 2391.478 | 44232.5398 | 4.5 | 2430.1369 | 3.5 | 2.97 | 0.19 | -1.59 | 0.03 |
| 2395.419 | 42401.3198 | 1.5 | 667.6829 | 2.5 | 25.1 | 1.1 | -1.06 | 0.02 |
| 2395.625 | 42114.8380 | 4.5 | 384.7872 | 3.5 | 266. | 4. | 0.36 | 0.01 |
| 2399.241 | 42334.8444 | 2.5 | 667.6829 | 2.5 | 137.3 | 3.3 | -0.15 | 0.01 |
| 2402.599 | 44446.9051 | 3.5 | 2837.9807 | 2.5 | 2.13 | 0.13 | -1.83 | 0.03 |
| 2404.431 | 42439.8511 | 0.5 | 862.6118 | 1.5 | 59. | 5. | -0.99 | 0.04 |
| 2404.885 | 42237.0575 | 3.5 | 667.6829 | 2.5 | 205. | 5. | 0.15 | 0.01 |
| 2406.661 | 42401.3198 | 1.5 | 862.6118 | 1.5 | 167. | 6. | -0.24 | 0.02 |
| 2410.519 | 42334.8444 | 2.5 | 862.6118 | 1.5 | 153.7 | 3.7 | -0.10 | 0.01 |
| 2411.067 | 42439.8511 | 0.5 | 977.0498 | 0.5 | 242. | 22. | -0.38 | 0.04 |
| 2413.310 | 42401.3198 | 1.5 | 977.0498 | 0.5 | 99.2 | 3.4 | -0.46 | 0.01 |
| 2451.101 | 42658.2444 | 3.5 | 1872.5998 | 4.5 | 0.324 | 0.020 | -2.63 | 0.03 |
| 2526.833 | 42401.3198 | 1.5 | 2837.9807 | 2.5 | 0.68 | 0.07 | -2.58 | 0.04 |
| 2562.535 | 46967.4751 | 2.5 | 7955.3186 | 3.5 | 184. | 6. | 0.03 | 0.01 |
| 2563.475 | 47389.809 | 1.5 | 8391.9554 | 2.5 | 145. | 5. | -0.25 | 0.01 |
| 2566.912 | 47626.110 | 0.5 | 8680.4706 | 1.5 | 117.2 | 3.8 | -0.64 | 0.01 |
| 2577.922 | 47626.110 | 0.5 | 8846.7837 | 0.5 | 126. | 4. | -0.60 | 0.01 |
| 2582.583 | 47389.809 | 1.5 | 8680.4706 | 1.5 | 84.9 | 2.7 | -0.47 | 0.01 |
| 2585.876 | 38660.0537 | 3.5 | 0.0000 | 4.5 | 87.2 | 2.1 | -0.16 | 0.01 |
| 2591.543 | 46967.4751 | 2.5 | 8391.9554 | 2.5 | 55.1 | 2.0 | -0.48 | 0.02 |
| 2593.728 | 47389.809 | 1.5 | 8846.7837 | 0.5 | 14.9 | 0.5 | -1.22 | 0.01 |
| 2598.369 | 38858.9696 | 2.5 | 384.7872 | 3.5 | 143.1 | 3.0 | -0.06 | 0.01 |
| 2599.395 | 38458.9934 | 4.5 | 0.0000 | 4.5 | 236. | 4. | 0.38 | 0.01 |
| 2607.087 | 39013.2160 | 1.5 | 667.6829 | 2.5 | 174. | 4. | -0.15 | 0.01 |
| 2611.073 | 46967.4751 | 2.5 | 8680.4706 | 1.5 | 7.23 | 0.31 | -1.36 | 0.02 |
| 2611.873 | 38660.0537 | 3.5 | 384.7872 | 3.5 | 124.9 | 3.0 | 0.01 | 0.01 |
| 2613.824 | 39109.3161 | 0.5 | 862.6118 | 1.5 | 215. | 6. | -0.36 | 0.01 |
| 2617.617 | 38858.9696 | 2.5 | 667.6829 | 2.5 | 48.5 | 1.0 | -0.52 | 0.01 |
| 2620.409 | 39013.2160 | 1.5 | 862.6118 | 1.5 | 3.44 | 0.13 | -1.85 | 0.02 |
| 2621.669 | 39109.3161 | 0.5 | 977.0498 | 0.5 | 53.6 | 1.6 | -0.96 | 0.01 |
| 2625.667 | 38458.9934 | 4.5 | 384.7872 | 3.5 | 34.9 | 1.2 | -0.44 | 0.01 |
| 2628.293 | 39013.2160 | 1.5 | 977.0498 | 0.5 | 87.0 | 2.6 | -0.44 | 0.01 |
| 2631.047 | 38858.9696 | 2.5 | 862.6118 | 1.5 | 81.6 | 1.7 | -0.29 | 0.01 |
| 2631.323 | 38660.0537 | 3.5 | 667.6829 | 2.5 | 59.8 | 1.8 | -0.30 | 0.01 |
| 2692.834 | 45079.9019 | 2.5 | 7955.3186 | 3.5 | 1.50 | 0.07 | -2.01 | 0.02 |
| 2714.413 | 44784.7859 | 2.5 | 7955.3186 | 3.5 | 54.8 | 1.4 | -0.44 | 0.01 |
| 2724.884 | 45079.9019 | 2.5 | 8391.9554 | 2.5 | 9.69 | 0.36 | -1.19 | 0.02 |
| 2727.539 | 45044.1916 | 1.5 | 8391.9554 | 2.5 | 98.0 | 3.2 | -0.36 | 0.01 |
| 2730.734 | 45289.8248 | 1.5 | 8680.4706 | 1.5 | 28.0 | 1.6 | -0.91 | 0.02 |
| 2732.448 | 38458.9934 | 4.5 | 1872.5998 | 4.5 | 0.076 | 0.006 | -3.07 | 0.03 |
| 2736.966 | 45206.4704 | 0.5 | 8680.4706 | 1.5 | 121. | 9. | -0.57 | 0.03 |
| 2739.547 | 44446.9051 | 3.5 | 7955.3186 | 3.5 | 237. | 5. | 0.33 | 0.01 |
| 2743.197 | 45289.8248 | 1.5 | 8846.7837 | 0.5 | 200. | 11. | -0.05 | 0.02 |
| 2746.483 | 45079.9019 | 2.5 | 8680.4706 | 1.5 | 216. | 7. | 0.16 | 0.01 |

| | | | | | | | | | |
|---|---|---|---|---|---|---|---|---|---|
| 2746.982 | 44784.7859 | 2.5 | 8391.9554 | 2.5 | 178. | 5. | 0.08 | 0.01 |
| 2749.181 | 45044.1916 | 1.5 | 8680.4706 | 1.5 | 129. | 4. | -0.24 | 0.01 |
| 2749.321 | 44753.8179 | 3.5 | 8391.9554 | 2.5 | 226. | 7. | 0.31 | 0.01 |
| 2749.486 | 45206.4704 | 0.5 | 8846.7837 | 0.5 | 117. | 8. | -0.58 | 0.03 |
| 2755.736 | 44232.5398 | 4.5 | 7955.3186 | 3.5 | 224. | 6. | 0.41 | 0.01 |
| 2761.813 | 45044.1916 | 1.5 | 8846.7837 | 0.5 | 13.7 | 0.5 | -1.20 | 0.02 |
| 2768.934 | 44784.7859 | 2.5 | 8680.4706 | 1.5 | 4.42 | 0.17 | -1.51 | 0.02 |
| 2868.874 | 43238.6066 | 2.5 | 8391.9554 | 2.5 | 0.85 | 0.08 | -2.21 | 0.04 |
| 2880.756 | 42658.2444 | 3.5 | 7955.3186 | 3.5 | 2.26 | 0.21 | -1.65 | 0.04 |
| 2917.466 | 42658.2444 | 3.5 | 8391.9554 | 2.5 | 0.179 | 0.023 | -2.74 | 0.05 |
| 2926.585 | 42114.8380 | 4.5 | 7955.3186 | 3.5 | 4.13 | 0.34 | -1.28 | 0.03 |
| 2939.507 | 42401.3198 | 1.5 | 8391.9554 | 2.5 | 0.40 | 0.05 | -2.68 | 0.05 |
| 2944.395 | 47626.110 | 0.5 | 13673.2045 | 1.5 | 55.1 | 2.8 | -0.85 | 0.02 |
| 2947.655 | 47389.809 | 1.5 | 13474.4474 | 2.5 | 30.2 | 1.8 | -0.81 | 0.03 |
| 2953.774 | 42237.0575 | 3.5 | 8391.9554 | 2.5 | 4.00 | 0.33 | -1.38 | 0.03 |
| 2961.275 | 42439.8511 | 0.5 | 8680.4706 | 1.5 | 0.90 | 0.17 | -2.63 | 0.08 |
| 2964.623 | 47626.110 | 0.5 | 13904.8604 | 0.5 | 9.5 | 0.8 | -1.60 | 0.04 |
| 2965.032 | 47389.809 | 1.5 | 13673.2045 | 1.5 | 8.6 | 0.6 | -1.35 | 0.03 |
| 2970.515 | 42334.8444 | 2.5 | 8680.4706 | 1.5 | 2.81 | 0.23 | -1.65 | 0.03 |
| 2975.936 | 42439.8511 | 0.5 | 8846.7837 | 0.5 | 0.98 | 0.16 | -2.59 | 0.07 |
| 2979.353 | 42401.3198 | 1.5 | 8846.7837 | 0.5 | 1.81 | 0.17 | -2.02 | 0.04 |
| 2984.825 | 46967.4751 | 2.5 | 13474.4474 | 2.5 | 44.9 | 2.6 | -0.45 | 0.02 |
| 2985.545 | 47389.809 | 1.5 | 13904.8604 | 0.5 | 24.0 | 1.6 | -0.89 | 0.03 |
| 3002.644 | 46967.4751 | 2.5 | 13673.2045 | 1.5 | 16.9 | 1.8 | -0.87 | 0.04 |
| 3170.337 | 45206.4704 | 0.5 | 13673.2045 | 1.5 | 1.30 | 0.23 | -2.38 | 0.07 |
| 3183.114 | 45079.9019 | 2.5 | 13673.2045 | 1.5 | 1.07 | 0.09 | -2.01 | 0.04 |
| 3186.737 | 45044.1916 | 1.5 | 13673.2045 | 1.5 | 3.84 | 0.26 | -1.64 | 0.03 |
| 3192.910 | 44784.7859 | 2.5 | 13474.4474 | 2.5 | 1.36 | 0.13 | -1.90 | 0.04 |
| 3193.801 | 45206.4704 | 0.5 | 13904.8604 | 0.5 | 7.4 | 0.7 | -1.65 | 0.04 |
| 3196.071 | 44753.8179 | 3.5 | 13474.4474 | 2.5 | 1.48 | 0.10 | -1.74 | 0.03 |
| 3210.445 | 45044.1916 | 1.5 | 13904.8604 | 0.5 | 3.80 | 0.29 | -1.62 | 0.03 |
| 3213.309 | 44784.7859 | 2.5 | 13673.2045 | 1.5 | 6.2 | 0.4 | -1.24 | 0.03 |
| 3227.743 | 44446.9051 | 3.5 | 13474.4474 | 2.5 | 7.5 | 0.5 | -1.03 | 0.03 |
| 3255.888 | 38660.0537 | 3.5 | 7955.3186 | 3.5 | 0.242 | 0.025 | -2.51 | 0.04 |
| 3277.349 | 38458.9934 | 4.5 | 7955.3186 | 3.5 | 0.45 | 0.09 | -2.14 | 0.08 |
| 4173.451 | 44784.7859 | 2.5 | 20830.5534 | 2.5 | 0.268 | 0.030 | -2.38 | 0.05 |
| 4233.162 | 44446.9051 | 3.5 | 20830.5534 | 2.5 | 0.45 | 0.05 | -2.02 | 0.05 |
| 4303.170 | 45044.1916 | 1.5 | 21812.0454 | 1.5 | 0.27 | 0.07 | -2.52 | 0.10 |
| 4351.762 | 44784.7859 | 2.5 | 21812.0454 | 1.5 | 0.66 | 0.11 | -1.95 | 0.07 |
| 4385.377 | 45206.4704 | 0.5 | 22409.8178 | 0.5 | 0.40 | 0.07 | -2.64 | 0.07 |
| 4416.819 | 45044.1916 | 1.5 | 22409.8178 | 0.5 | 0.23 | 0.05 | -2.57 | 0.09 |
| 4508.281 | 45206.4704 | 0.5 | 23031.2829 | 1.5 | 0.63 | 0.10 | -2.42 | 0.06 |
| 4522.628 | 45044.1916 | 1.5 | 22939.3512 | 2.5 | 0.42 | 0.06 | -2.29 | 0.06 |
| 4549.466 | 44784.7859 | 2.5 | 22810.3459 | 3.5 | 0.64 | 0.10 | -1.92 | 0.06 |
| 4583.829 | 44446.9051 | 3.5 | 22637.1950 | 4.5 | 0.45 | 0.06 | -1.94 | 0.05 |

Table 7
Iron Abundances in the Sun and HD 84937

| wavelength (Å) | χ (eV) | log(*gf*) this study | log *ε* Sun | log *ε* HD 84937 |
|---|---|---|---|---|
| 2327.396 | 0.083 | -0.68 | ... | 5.31 |
| 2331.308 | 0.232 | -0.75 | ... | 5.33 |
| 2332.799 | 0.048 | -0.25 | ... | 5.17 |
| 2343.495 | 0.000 | 0.06 | ... | 5.25 |
| 2354.890 | 0.352 | -1.07 | ... | 5.12 |
| 2359.107 | 0.107 | -0.66 | ... | 5.29 |
| 2360.294 | 0.301 | -0.59 | ... | 5.29 |
| 2368.596 | 0.352 | -0.69 | ... | 5.29 |
| 2379.276 | 0.301 | -0.83 | ... | 5.33 |
| 2380.761 | 0.083 | -0.68 | ... | 5.21 |
| 2385.006 | 0.352 | -1.70 | ... | 5.29 |
| 2391.478 | 0.301 | -1.59 | ... | 5.32 |
| 2526.833 | 0.352 | -2.58 | ... | 5.35 |
| 2563.475 | 1.040 | -0.25 | ... | 5.27 |
| 2566.912 | 1.076 | -0.64 | ... | 5.27 |
| 2577.922 | 1.097 | -0.60 | ... | 5.30 |
| 2582.583 | 1.076 | -0.47 | ... | 5.30 |
| 2585.876 | 0.000 | -0.16 | ... | 5.31 |
| 2591.543 | 1.040 | -0.48 | ... | 5.23 |
| 2598.369 | 0.048 | -0.06 | ... | 5.22 |
| 2599.395 | 0.000 | 0.38 | ... | 5.25 |
| 2607.087 | 0.083 | -0.15 | ... | 5.24 |
| 2611.073 | 1.076 | -1.36 | ... | 5.27 |
| 2611.873 | 0.048 | 0.01 | ... | 5.21 |
| 2613.824 | 0.107 | -0.36 | ... | 5.24 |
| 2617.617 | 0.083 | -0.52 | ... | 5.10 |
| 2620.408 | 0.107 | -1.85 | ... | 5.35 |
| 2621.669 | 0.121 | -0.96 | ... | 5.32 |
| 2625.667 | 0.048 | -0.44 | ... | 5.25 |
| 2628.293 | 0.121 | -0.44 | ... | 5.27 |
| 2692.834 | 0.986 | -2.01 | ... | 5.27 |
| 2714.413 | 0.986 | -0.44 | ... | 5.30 |
| 2724.884 | 1.040 | -1.19 | ... | 5.17 |
| 2730.734 | 1.076 | -0.91 | ... | 5.30 |
| 2732.448 | 0.232 | -3.07 | ... | 5.29 |
| 2739.547 | 0.986 | 0.33 | ... | 5.29 |
| 2743.197 | 1.097 | -0.05 | ... | 5.17 |
| 2746.483 | 1.076 | 0.16 | ... | 5.22 |
| 2746.982 | 1.040 | 0.08 | ... | 5.25 |
| 2755.736 | 0.986 | 0.41 | ... | 5.20 |
| 2761.813 | 1.097 | -1.20 | ... | 5.27 |
| 2768.934 | 1.076 | -1.51 | ... | 5.25 |
| 2880.756 | 0.986 | -1.65 | ... | 5.27 |
| 2917.462 | 1.040 | -2.74 | ... | 5.32 |
| 2926.585 | 0.986 | -1.28 | ... | 5.35 |
| 2939.507 | 1.040 | -2.68 | ... | 5.37 |
| 2944.395 | 1.695 | -0.85 | ... | 5.21 |
| 2947.655 | 1.671 | -0.81 | ... | 5.27 |
| 2961.273 | 1.076 | -2.63 | ... | 5.47 |
| 2964.624 | 1.724 | -1.60 | ... | 5.17 |
| 2965.032 | 1.695 | -1.35 | ... | 5.24 |
| 2970.515 | 1.076 | -1.65 | ... | 5.22 |
| 2975.933 | 1.097 | -2.59 | ... | 5.27 |
| 2979.353 | 1.097 | -2.02 | ... | 5.28 |
| 2984.825 | 1.671 | -0.45 | ... | 5.19 |
| 2985.545 | 1.724 | -0.89 | ... | 5.24 |
| 3002.643 | 1.695 | -0.87 | ... | 5.19 |
| 3170.340 | 1.695 | -2.38 | ... | 5.09 |
| 3183.114 | 1.695 | -2.01 | ... | 5.17 |
| 3186.737 | 1.695 | -1.64 | ... | 5.23 |
| 3192.910 | 1.671 | -1.90 | ... | 5.24 |
| 3193.801 | 1.724 | -1.65 | ... | 5.24 |
| 3196.071 | 1.671 | -1.74 | ... | 5.24 |
| 3255.888 | 0.986 | -2.51 | ... | 5.25 |
| 3277.349 | 0.986 | -2.14 | ... | 5.07 |
| 4173.450 | 2.583 | -2.38 | 7.45 | 5.07 |
| 4233.163 | 2.583 | -2.02 | 7.50 | 5.32 |
| 4303.168 | 2.704 | -2.52 | ... | 5.30 |
| 4385.379 | 2.778 | -2.64 | 7.50 | 5.32 |
| 4416.818 | 2.778 | -2.57 | 7.40 | 5.24 |
| 4508.281 | 2.856 | -2.42 | 7.50 | 5.34 |
| 4522.628 | 2.844 | -2.29 | 7.55 | 5.38 |
| 4549.467 | 2.828 | -1.92 | 7.30 | 5.17 |
| 4583.829 | 2.807 | -1.94 | 7.50 | 5.34 |

| ALL LINES | Sun | HD 84937 |
|---|---|---|
| average | 7.46 | 5.26 |
| p/m | 0.03 | 0.01 |
| sigma | 0.08 | 0.07 |
| count | 8 | 74 |

| λ>4100 Å | | |
|---|---|---|
| average | 7.46 | 5.28 |
| p/m | 0.03 | 0.03 |
| sigma | 0.08 | 0.10 |
| count | 8 | 9 |